\documentclass[letterpaper,twocolumn,10pt]{article}
\usepackage[dvipsnames,table]{xcolor}
\usepackage{usenix2019_v3}
\usepackage{tikz}
\usepackage{subfigure}
\usepackage{multirow}
\usepackage{amsmath}
\usepackage{lipsum}
\usepackage{threeparttable}
\usepackage{makecell}
\usepackage{titlesec}
\usepackage[ruled,vlined, linesnumbered]{algorithm2e}
\usepackage{algorithmic}
\usepackage{listings}
\usepackage{code_styles/diff/lang}
\usepackage{code_styles/diff/style}
\usepackage[strict]{changepage}
\usepackage{enumitem}
\usepackage{float}
\usepackage{lipsum}
\usepackage{lmodern}
\usepackage[most]{tcolorbox}
\usepackage{mathptmx}
\usepackage{amsfonts}
\usepackage{pifont}
\usepackage{CJKutf8}
\usepackage{newtxtext,newtxmath}
\usepackage{tabularray}
\usepackage{booktabs}
\usepackage{graphicx}
\usepackage[numbers,sort&compress]{natbib}
\usepackage{hyperref}
\usepackage{cleveref}
\usepackage{url}
\usepackage{wasysym}

\usepackage{bbding}
\usepackage{ulem}
\usepackage{diagbox}
\usepackage{threeparttable}
\usepackage{caption}
\usepackage{enumitem}
\usepackage{appendix}
\usepackage{stfloats}
\usepackage{setspace}
\usepackage{authblk}

\newcommand{\newtext}[1]{\textcolor{black} {#1}}

\newcommand{\subject}[1]{\vspace{3pt}\noindent\textbf{#1}}
\newcommand{\subsubject}[1]{\vspace{3pt}\textit{#1}}

\newcommand*\emptycirc[1][1ex]{\tikz\draw (0,0) circle (#1);} 
\newcommand*\halfcirc[1][1ex]{%
	\begin{tikzpicture}
	\draw[fill] (0,0)-- (90:#1) arc (90:270:#1) -- cycle ;
	\draw (0,0) circle (#1);
	\end{tikzpicture}}
\newcommand*\fullcirc[1][1ex]{\tikz\fill (0,0) circle (#1);}

\begin{document}
%-------------------------------------------------------------------------------

\title{SoK: Towards Effective Automated Vulnerability Repair}

\author[1]{Ying Li}
\author[2]{Faysal Hossain Shezan}
\author[1]{Bomin Wei}
\author[3]{Gang Wang}
\author[1]{Yuan Tian}
\affil[1]{UCLA}
\affil[2]{University of Texas at Arlington}
\affil[3]{University of Illinois Urbana-Champaign}
\affil[ ]{\textit{\{yinglee,davidwei23,yuant\}@ucla.edu,
faysal.shezan@uta.edu,
gangw@illinois.edu}}

% \author{
%     {\rm Your N.\ Here} (Your Institution)
%     \and
%     {\rm Second Name} (Second Institution)
%     \and
%     {\rm Third Name} (Third Institution)
% }

% \author{
% {\rm Ying Li}\\
% Your Institution
% \and
% {\rm Second Name}\\
% Second Institution
% \and
% {\rm Second Name}\\
% Second Institution
% % copy the following lines to add more authors
% \and
% {\rm Name}\\
% Name Institution
% } % end author

\pagestyle{plain}

\maketitle
\thispagestyle{plain}

\newcommand{\revised}[2]{
    \textcolor{red}{\sout{#1}}\textcolor{blue}{#2}
}

\newcounter{definition}

\newenvironment{definition}%
{%
    \refstepcounter{definition}% 
    \textbf{\textit{Definition} \thedefinition. }\space % 
}%
{%
}
%\renewcommand{\thefootnote}{\fnsymbol{footnote}}
%\footnotetext[1]{Corresponding author.}
\newcounter{problem}

\newenvironment{problem}%
{%
    \refstepcounter{problem}% 
    \textbf{Problem Definition.}\space% 
}%
{%
}
%
%\newcommand\malurl[1]{\href{notalink}{\textcolor{blue}{\nolinkurl{#1}}}}
% \newcommand\malurl[1]{\href{notalink}{{\nolinkurl{#1}}}}
% % the finding box
% \newcounter{finding}
% \setcounter{finding}{1}
% \newcommand{\finding}[1]{
% \vspace{3pt}
% \noindent
% \framebox{
% \begin{minipage}[b]{.95\columnwidth}
% \noindent \textbf{Finding \Roman{finding}}: {#1} %\textit{#1}
% \stepcounter{finding}
% \end{minipage}
% }
% % \vspace{-5pt}
% }

\newcounter{finding}
\setcounter{finding}{1}
\newcommand{\finding}[1]{
\vspace{1pt}
\noindent
\begin{tcolorbox}[ enhanced, 
    breakable,
    boxrule=0.5pt,
    arc=4pt,
    left=2pt,
    right=2pt,
    bottom=2pt,
    top=2pt,
    rounded corners]
\noindent \textbf{Finding \Roman{finding}.} \small
\textit{#1}
\stepcounter{finding}
\end{tcolorbox}
% \vspace{-5pt}
}

% \newcounter{findingCounter}

% \newcommand{\finding}[1]{
%   \begin{tcolorbox}[enhanced, left=3mm,right=3mm,
%     colback=gray!10, colframe=gray!80, boxrule=0pt,
%     borderline west={4pt}{0pt}{gray!90},
%     breakable
%     ]
    
%     \textbf{Finding \Roman{findingCounter}}: {#1}
%     \stepcounter{findingCounter}
%     \end{tcolorbox}
% }

% \newcounter{openrq}
% \setcounter{openrq}{1}
% \newcommand{\openrq}[1]{
% \vspace{3pt}
% \noindent
% \framebox{
% \begin{minipage}[b]{.95\columnwidth}
% \noindent \textbf{Open RQ \Roman{openrq}}: {#1} %\textit{#1}
% \stepcounter{openrq}
% \end{minipage}
% }
% % \vspace{-5pt}
% }

\newcounter{openrq}
\setcounter{openrq}{1}
\newcommand{\openrq}[1]{
\vspace{1pt}
\noindent
\begin{tcolorbox}[ enhanced, 
    breakable,
    boxrule=0.5pt,
    arc=4pt,
    left=2pt,
    right=2pt,
    bottom=2pt,
    top=2pt,
    rounded corners]
\noindent \textbf{Open RQ \Roman{openrq}.} \small
\textit{#1}
\stepcounter{openrq}
\end{tcolorbox}
% \vspace{-5pt}
}

\newcounter{takeaway}
\setcounter{takeaway}{1}
\newcommand{\takeaways}[1]{
\vspace{1pt}
\noindent
\begin{tcolorbox}[ enhanced, 
    breakable,
    boxrule=0.5pt,
    arc=4pt,
    left=2pt,
    right=2pt,
    bottom=2pt,
    top=2pt,
    rounded corners]
\noindent \textbf{Takeaway \Roman{takeaway}.} \small
\textit{#1}
\stepcounter{takeaway}
\end{tcolorbox}
% \vspace{-5pt}
}

% \newcounter{takeaway}
% \setcounter{takeaway}{1}
% \newcommand{\takeaways}[1]{
% \vspace{1pt}
% \noindent
% \begin{tcolorbox}[boxrule=0.5pt]
% \noindent \textbf{Takeaway \Roman{takeaway}.}
% \textit{#1}
% \stepcounter{takeaway}
% \end{tcolorbox}
% % \vspace{-5pt}
% }

\newcounter{oproblem}
\setcounter{oproblem}{1}
\newcommand{\openquestion}[1]{
\vspace{1pt}
\noindent
\begin{tcolorbox}[ enhanced, 
    breakable,
    boxrule=0.5pt,
    arc=4pt,
    left=2pt,
    right=2pt,
    bottom=2pt,
    top=2pt,
    rounded corners]
\noindent \textbf{Open Problems \Roman{oproblem}.}\small
\textit{#1}
\stepcounter{oproblem}
\end{tcolorbox}
% \vspace{-5pt}
}

\newcommand{\recprac}[1]{
\vspace{1pt}
\noindent
\begin{tcolorbox}[enhanced, 
    breakable,
    boxrule=0.5pt,
    arc=4pt,
    left=2pt,
    right=2pt,
    bottom=2pt,
    top=2pt,
    rounded corners]
\noindent \textbf{Recommended Practices.} \small
\textit{#1}
\end{tcolorbox}
% \vspace{-5pt}
}
\newcommand{\para}[1]{{\vspace{2pt} \bf \noindent #1 \hspace{1pt}}}

\newcommand{\ignore}[1]{}

\newenvironment{takeawaynew}[1][]
  {
    % \vspace{-0.1em}
 \begin{tcolorbox}
 [%
    enhanced, 
    breakable,
    boxrule=0.5pt,
    arc=4pt,
    left=2pt,
    right=2pt,
    bottom=2pt,
    top=2pt,
    % colback=boxcolor, 
    % colframe=black,
    rounded corners
    % frame hidden,
    % overlay broken = {
    %     \draw[line width=0pt, black, rounded corners]
    %     (frame.north west) rectangle (frame.south east);},
    ]{}
  \textbf{#1.}
  \small \itshape}
  {
\end{tcolorbox}
    % \vspace{-0.1em}
}
\newcommand{\papernumber}{79\xspace} %68
\newcommand{\manualcnt}{56\xspace}
\newcommand{\manualprop}{70.89\%\xspace}
\newcommand{\sourcelevelno}{63\xspace}
\newcommand{\benchmarkno}{\xspace}
\newcommand{\binarylevelno}{5\xspace}
\newcommand{\templatecnt}{29\xspace} 
\newcommand{\nongncnt}{31\xspace} 
\newcommand{\Anumber}{67\xspace}

\newcommand{\nongncppcnt}{16\xspace} 
\newcommand{\memorycnt}{10\xspace}

\newcommand{\arxivcnt}{8\xspace}
\newcommand{\approachcnt}{10\xspace}
\newcommand{\chkappa}{91.5\%\xspace}
\newcommand{\agreerate}{96.5\%\xspace}

\newcommand{\artifactcnt}{32\xspace}
\newcommand{\srccnt}{71\xspace}
\newcommand{\bincnt}{4\xspace}
\newcommand{\btcnt}{3\xspace}

\newcommand{\cvecnt}{126\xspace}
\newcommand{\cwecnt}{98\xspace}

\begin{abstract}

The increasing prevalence of software vulnerabilities necessitates automated vulnerability repair (AVR) techniques. This Systematization of Knowledge (SoK) provides a comprehensive overview of the AVR landscape, encompassing both synthetic and real-world vulnerabilities. Through a systematic literature review and quantitative benchmarking across diverse datasets, methods, and strategies, we establish a taxonomy of existing AVR methodologies, categorizing them into template-guided, search-based, constraint-based, and learning-driven approaches. We evaluate the strengths and limitations of these approaches, highlighting common challenges and practical implications. Our comprehensive analysis of existing AVR methods 
reveals a diverse landscape with no single ``best'' approach. Learning-based methods excel in specific scenarios but lack complete program understanding, and both learning and non-learning methods face challenges with complex vulnerabilities. 
Additionally, we identify emerging trends and propose future research directions to advance the field of AVR. This SoK serves as a valuable resource for researchers and practitioners, offering a structured understanding of the current state-of-the-art and guiding future research and development in this critical domain.

\end{abstract}

%-------------------------------------------------------------------------------
% Section
%-------------------------------------------------------------------------------

\section{Introduction}
\label{sec:introduction}

The relentless increase in software vulnerabilities poses a critical challenge for organizations, with consequences ranging from financial loss to reputational damage~\cite{vul_consequence, shin2010evaluating}. This challenge is compounded by the limitations of manual repair processes, which are often slow, error-prone, and struggle to keep up with the volume of necessary fixes.  Furthermore, numerous known vulnerabilities remain unpatched for extended periods, leaving systems exposed~\citep{li2017large, li2017novel,vul_remidition_timeline}. Automatic Vulnerability Repair (AVR) has emerged as a vital field to address these challenges, offering the potential to significantly reduce the time and resources to mitigate security risks.
% }

AVR has witnessed significant progress, evolving from the template-based approach that applied predefined repairs (often inflexible), random mutation, to more sophisticated approaches~\cite{lin2007autopag, novark2007exterminator, coker2013program}. For example, constraint-solving techniques, often deriving constraints from symbolic execution, can be computationally expensive~\cite{gao2021beyond, zhang2022program}. Deep learning offers an end-to-end translation approach, but its effectiveness hinges on high-quality vulnerability repair datasets~\cite{goodfellow2020generative, chi2022seqtrans, chen2022neural}. Currently, large language models (LLMs) show promise due to their vast training data, although limitations persist in function-level repairs and comprehensive program understanding~\cite{wu2023effective, pearce2023examining, fu2023chatgpt}.

Given the rapid growth of this field, a \newtext{SoK} is crucial to understand the landscape. This SoK aims to:
(1) Bridge knowledge gaps by analyzing AVR approaches across synthetic data and real-world vulnerabilities.
(2) Offer critical insights into the strengths and limitations of current work.
(3) Propose promising future research directions.
In this paper, we establish a taxonomy of existing AVR methodologies, grounded in fundamental principles, \newtext{and conduct a quantitative comparison of current approaches.}
This SoK provides a comparative analysis of security patch generation methods, complexity levels, impact factors for successful repair, and vulnerability types, revealing a multifaceted landscape. We assess both synthetic and real-world vulnerabilities, finding no definitive ``best'' method; each has inherent limitations. For example, learning-based methods lack whole-program understanding, while non-learning methods struggle with precise constraint extraction, e.g., loop invariant synthesis. 
Based on our taxonomy and benchmarks, we analyze current research advancements, theoretical challenges, and future directions in AVR. Our findings indicate that performance across existing works varies significantly depending on the benchmark used, which often suffer from limited data points. 
We propose several compelling research directions to advance AVR, including the integration of hybrid approaches, \newtext{interpretability of AVR}, and the generation of high-quality specifications.

This SoK benefits both experts and practitioners. For security engineers, it provides (1) a systematic taxonomy to understand existing research; (2) analysis for leading repair methods; and (3) insights into limitations and future potential. Practitioners gain (1) a concise problem definition of vulnerability repair and (2) a comprehensive evaluation with practical implementations of key approaches.
% }
In summary, we have the following key contributions:
\begin{itemize}[nosep, leftmargin=*]
    \item We define the core problem of security vulnerability repair and provide a comprehensive taxonomy and comparative analysis of repair approaches (Section~\ref{sec:taxonomy_apr}), with a focus on patch generation techniques (Section~\ref{sec:patch_generation}). This discussion culminates in actionable takeaways and clearly identified open research questions.
     \item Our taxonomy classifies security patch generation approaches by 4 categories, 5+ strategies, 11+ methodologies, \newtext{5+} programming languages, and \newtext{12+} vulnerability types.
    \item We conduct quantitative evaluations of existing methods across C/C++ and Java benchmarks. We synthesize practical implications and highlight prevalent challenges in AVR, offering a critical assessment of the field's progress.
    \item  We explore the latest research trends, identify ongoing challenges and propose directions for future work. Through quantitative and qualitative analysis, we highlight gaps in current techniques and extract key findings to guide future advancements in AVR. 
    We will keep our website (\url{https://sok-avr.github.io}) updated with all the latest studies and findings to promote the research in this field.
\end{itemize}

\subject{Related Work.}  
\newtext{This study mainly differs from existing surveys in the following three aspects: (i) Existing works~\citep{monperrus2018automatic, gazzola2018automatic, monperrus2018living,pinconschi2021comparative, winter2022let, huang2023survey, zhang2023survey} focus on program repair from a general perspective rather than a security-specific one. For instance, Huang et al.\cite{huang2023survey} briefly discusses security vulnerabilities in the context of learning-based methods, Monperrus et al.\cite{monperrus2018living} include only a subset of AVR-related papers, and Pinconschi et al.~\cite{pinconschi2021comparative} demonstrated that general program repair tools perform poorly on vulnerability repair. These gaps highlight the need for our security-focused analysis. 
(ii) existing surveys on vulnerability repair focuses on LLM for AVR~\citep{wu2023effective, zhou2024large}, however, traditional methods (e.g., constraint-based methods) for AVR continue to evolve and demonstrate effectiveness in recent research~\citep{xing8if,yu2024tapfixer}; (iii) Prior surveys lack a systematic comparison across different AVR approaches. Our study bridges these gaps by providing both quantitative benchmarking across diverse datasets and qualitative analysis of various approaches, enabling researchers to better understand the current landscape and identify promising future research directions in AVR.}

\section{Preliminaries and Problem Setup}
\label{sec: preliminaries}

\begin{figure}
    \centering
    \includegraphics[width=0.5\textwidth]{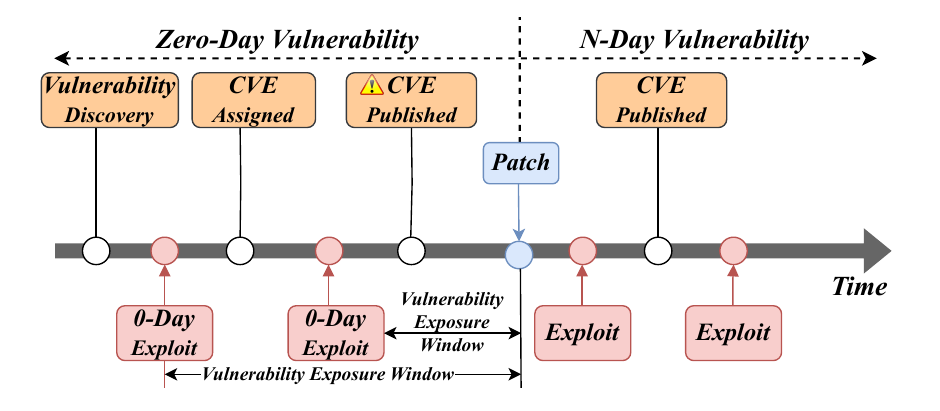}
    \caption{The timeline of vulnerability discovery, patch release, and exploit publication. 
    }
    \label{fig:vul_cycle}
    \vspace{-1em}
\end{figure}

 Figure~\ref{fig:vul_cycle} shows the timeline from vulnerability discovery to CVE publication and vulnerability repair process. The timeline from vulnerability discovery to CVE publication typically involves several key stages. Initially, the exact discovery time of a vulnerability is often unknown. CVEs are usually published after a vendor releases a patch, but in urgent cases or due to lack of cooperation, they may be disclosed before a patch is available. A vulnerability is classified as a zero-day until a patch is released, after which it becomes an N-day vulnerability. The main goal of AVR is to minimize the vulnerability exposure window. 
\newtext{The vulnerability repair workflow includes three key components: \textit{vulnerability localization} (VL), \textit{security patch generation} (SPG), and \textit{security patch validation} (SPV). Below we detail each repair stage and further illustrate its relationship to program repair.}

\subsection{Vulnerability Localization}
VL identifies the precise location or source of vulnerabilities within an application. This process involves pinpointing specific program points (i.e., specific locations within a program, e.g., lines of code, statements) where vulnerabilities exist and can be exploited, and then finding the point at which the ``root cause'' of the bug can be fixed~\citep{shen2021localizing}.

\begin{definition}
    % \textit{Vulnerability Localization.}
     We denote ${P}$ as an original vulnerable program. Given ${V}$ as the set of identified vulnerabilities within ${P}$, each vulnerability $v \in {V}$ is associated with a subset of program points {\small ${PT}_v\subseteq {PT}$}, where ${PT}$ is the set of all program points in $P$. The vulnerability location $L$ is the exact program point where the vulnerability $v$ manifests can be fixed.
\end{definition}

\newtext{Accurately locating vulnerabilities is crucial for AVR, as crash points often differ from actual vulnerability locations, which are actionable for patching~\cite{shen2021localizing}. Early methods used slice-based VL: (1) static slicing analyzes control and data flow dependencies through dependency graphs but often includes irrelevant statements~\cite{mirsky2023vulchecker}; (2) dynamic slicing~\cite{agrawal1990dynamic} records execution-specific data and control flow to extract relevant statements for given inputs~\cite{rajput2023icspatch}. Further, program state-based methods (e.g., delta debugging~\cite{zeller2002simplifying}) analyze runtime state changes by comparing states from successful and failed executions to pinpoint root causes.
However, these methods are computationally expensive. In contrast, spectrum-based methods efficiently locate faults using only test coverage information by analyzing statement execution patterns in failed/passed tests and calculating suspiciousness scores~\cite{xu2024racing}.} 

\newtext{Statistical reasoning like ConcFuzz~\cite{shen2021localizing}, combines fuzzing near exploit paths with statistical analysis for precise VL. Building on this, learning-based methods further improve detection. For example, Xu et al.~\cite{xu2024racing} locates vulnerabilities using reinforcement learning-guided fuzzing to generate counterexamples, combined with spectrum-based methods for VL. Recently, LLMs have shown potential in VL, Zhang et al.~\cite{zhang2024empirical} transforms the VL task into a sequence labeling problem, where each line of code is classified as either ``vulnerable'' or ``non-vulnerable'', which has achieved 82.0\% F1 score.}

\subsection{Security Patch Generation}
A security patch is created to fix existing vulnerabilities, with the goal of making the program secure after applying it~\cite{wang2023graphspd}. In contrast, non-security patches address issues like functionality bugs, feature updates or performance improvements. SPG is the process of creating critical patches for vulnerable code segments to effectively repair existing vulnerabilities.

\begin{definition}
    Given a vulnerable program ${P}$, with identified vulnerabilities $V$ at the specific locations $L$, a security patch $P'$ is a modified version of $P$.
    The process of generating ${P'}$ involves modifying $P$ at each location $l$ in $L$ to fix the vulnerabilities present at that location.
    
\end{definition}

\subsection{Security Patch Validation}
\label{subsec:sec_val}
After generating security patches, the pivotal step is to verify the if the patch fixes the given vulnerability without introducing any new bugs~\citep{bohme2017directed, jiang2022evocatio} while maintaining correct program behavior~\citep{chen2021fast, gazzola2017automatic}. We define SPV as:

\begin{definition}
    $SPV(P,P')$ evaluates if $P'$ meets the following criteria: (i) fix the given vulnerability in $P$; (ii) ensure no new bugs or vulnerabilities are introduced; (iii) preserve the original functionality of $P$. If all these three conditions are satisfied, the patch is deem to be an effective security fix.
\end{definition}

\newtext{Li et al.~\cite{li2017large} found approximately 12\% of security patches are defective - with 7\% fail to fully fix the target vulnerability, while 5\% introduce new ones. The validation employs three main approaches: (1) Static analysis using code features~\citep{xin2017leveraging, le2017s3}, expert rules~\cite{tan2016anti}, etc. to assess patch correctness without execution. For example, S3~\cite{le2017s3} prioritizes patches by analyzing six static features measuring syntactic and semantic differences from buggy code; (2) Dynamic testing through test cases/exploits~\citep{jin2010automated, xin2017identifying, taneja2008diffgen, kim2013automatic}, execution traces~\cite{xiong2018identifying} to test the patched code for verifying expected behavior; and (3) Formal methods~\citep{kim2023patchverif, gallagher2014verifying} like PATCHVERIF~\cite{kim2023patchverif} used symbolic execution and physical invariant checks verify whether patch-introduced behavioral modifications meet expectations.
Nowadays, LLMs like LLM4PatchCorrect~\cite{zhou2024leveraging} have been applied to predict patch correctness using contextual information from execution traces, bug descriptions.
Despite this, human effort is still required in some AVR works (See in Section~\ref{sec:taxonomy_apr}).}

\subsection{Security Vulnerability Repair}

\begin{problem}
    Security Vulnerability Repair (SVR) involves generating $P'$ by modifying a vulerable program $P$ at identified locations $L$ to fix vulnerabilities $V$, ensuring that $P'$ satisfies the correctness criteria $C$. The objective is to find an appropriate repair function $F$ that meets these requirements and $SPV(P,P')==True$.
\end{problem}

\newtext{The three key statges of SVR (i.e., VL, SPG, and SPV) work in tandem to ensure effective SVR. VL identifies the exact program points $L$ of vunlnerabilities $V$, providing inputs for SPG to generate patches that transform the vulnerable program $P$ into repaired program $P'$. SPV verifies that $P'$ fixes the vulnerabilities, preserves functionality, and introduces no new issues. These stages are interdependent, as errors in localization can propagate through SPG, while SPV feedback guides earlier steps, ensuring an iterative and reliable SVR.}

 \vspace{-2em}
\newtext{\subsection{Automatic Vulnerability Repair v.s. Automatic Program Repair}}

\newtext{AVR is a subset of Automatic Program Repair (APR). While both aim to automate the repair process and share fundamental stages—localization, patch generation, and patch validation—they differ significantly in principles, assumptions, and input settings, particularly during the patch generation phase, which introduces unique challenges for AVR.}

\newtext{APR primarily targets general software issues, such as functional bugs~\cite{jiang2021cure, yin2024thinkrepair} and compilation errors~\cite{ahmed2018compilation, mesbah2019deepdelta, li2022transrepair}. In contrast, AVR focuses on vulnerabilities, requiring patches that both eliminate exploitable behaviors and preserve the program's functional invariants. This dual requirement introduces additional complexity to the repair synthesis.}

\newtext{APR and AVR also differ in their assumptions. Traditional APR assumes program correctness can be fully captured through comprehensive test cases, while AVR relies on specific security properties and exploit conditions. Despite this difference, both approaches share the belief that effective repairs can be derived from patterns observed in historical fixes, enabling learning-based approaches in both fields.}

\newtext{The distinction in assumptions is most evident in their input requirements. APR typically relies on test suites to guide the repair process, while AVR operates with a single Proof-of-Concept (PoC) exploit that demonstrates the vulnerability. This limited information introduces unique technical challenges for AVR, as it must simultaneously ensure vulnerability elimination and functionality preservation without comprehensive test coverage.}

\begin{table*}[!ht]
\caption{Properties and References of Security Patch Generation Approaches. }
\label{tab:gen_taxonomy} 
\begin{threeparttable}
\centering
\resizebox*{\textwidth}{!}{
\SetTblrInner{rowsep=0pt}
\begin{tblr}{
  % colspec={Q[c]Q[c]Q[c]Q[c]Q[c]Q[c]P{1cm}Q[c]Q[c]},
  % colspec = {Q[c] Q[c] Q[c] Q[c] Q[c] Q[c] Q[c, p{4cm}] Q[c] Q[c]},
  row{1} = {c},
  row{2} = {c},
  row{3} = {c},
  row{4} = {c},
  row{5} = {c},
  row{6} = {c},
  row{7} = {c},
  row{8} = {c},
  row{9} = {c},
  row{10} = {c},
  row{11} = {c},
  row{12} = {c},
  row{13} = {c},
  row{15} = {c},
  row{16} = {c},
  row{17} = {c},
  row{18} = {c},
  row{21} = {c},
  row{22} = {c},
  row{23} = {c},
  row{24} = {c},
  row{25} = {c},
  row{26} = {c},
  row{27} = {c},
  row{28} = {c},
  row{29} = {c},
  row{30} = {c},
  row{31} = {c},
  row{32} = {c},
  row{34} = {c},
  row{35} = {c},
  row{36} = {c},
  row{37} = {c},
  row{38} = {c},
  row{39} = {c},
  row{40} = {c},
  row{41} = {c},
  row{42} = {c},
  row{43} = {c},
  row{44} = {c},
  row{46} = {c},
  row{47} = {c},
  row{48} = {c},
  row{49} = {c},
  row{50} = {c},
  cell{2}{1} = {r=17}{},
  cell{14}{4} = {c},
  cell{14}{5} = {c},
  cell{14}{6} = {c},
  cell{14}{7} = {c},
  cell{14}{8} = {c},
  cell{14}{9} = {c},
  cell{19}{2} = {c},
  cell{19}{3} = {c},
  cell{19}{4} = {c},
  cell{19}{5} = {c},
  cell{19}{6} = {c},
  cell{19}{7} = {c},
  cell{19}{8} = {c},
  cell{19}{9} = {c},
  cell{20}{1} = {r=13}{},
  cell{20}{2} = {c},
  cell{20}{3} = {c},
  cell{20}{4} = {c},
  cell{20}{5} = {c},
  cell{20}{6} = {c},
  cell{20}{7} = {c},
  cell{20}{8} = {c},
  cell{20}{9} = {c},
  cell{33}{1} = {r=14}{},
  cell{33}{2} = {c},
  cell{33}{3} = {c},
  cell{33}{4} = {c},
  cell{33}{5} = {c},
  cell{33}{6} = {c},
  cell{33}{7} = {c},
  cell{33}{8} = {c},
  cell{33}{9} = {c},
  cell{45}{3} = {c},
  cell{45}{4} = {c},
  cell{45}{5} = {c},
  cell{45}{6} = {c},
  cell{45}{7} = {c},
  cell{45}{8} = {c},
  cell{45}{9} = {c},
  cell{47}{1} = {r=9}{},
  cell{47}{2} = {r=2}{},
  cell{47}{3} = {r=2}{},
  cell{49}{2} = {r=2}{},
  cell{51}{3} = {c},
  cell{51}{4} = {c},
  cell{51}{5} = {c},
  cell{51}{6} = {c},
  cell{51}{7} = {c},
  cell{51}{8} = {c},
  cell{51}{9} = {c},
  cell{52}{3} = {c},
  cell{52}{4} = {c},
  cell{52}{5} = {c},
  cell{52}{6} = {c},
  cell{52}{7} = {c},
  cell{52}{8} = {c},
  cell{52}{9} = {c},
  cell{53}{3} = {c},
  cell{53}{4} = {c},
  cell{53}{5} = {c},
  cell{53}{6} = {c},
  cell{53}{7} = {c},
  cell{53}{8} = {c},
  cell{53}{9} = {c},
  cell{54}{3} = {c},
  cell{54}{4} = {c},
  cell{54}{5} = {c},
  cell{54}{6} = {c},
  cell{54}{7} = {c},
  cell{54}{8} = {c},
  cell{54}{9} = {c},
  cell{55}{3} = {c},
  cell{55}{4} = {c},
  cell{55}{5} = {c},
  cell{55}{6} = {c},
  cell{55}{7} = {c},
  cell{55}{8} = {c},
  cell{55}{9} = {c},
  vline{1-10} = {1-55}{},
  vline{2 -9} = {1-53}{},
  vline{3-9} = {3-18,20-33,35-46}{},
  vline{2,5-9} = {48}{},
  vline{2,4-9} = {50}{},
  hline{1} = {-}{},
  hline{56} = {-}{},
  hline{2,20, 33, 47} = {-}{},
  hline{3-14,16-18,19,21,23-25,27-28,30-32,33-37,39-41,43,45-46,48} = {4-9}{},
  hline{15,22,29,38,42,44,50,52-53,55} = {3-9}{},
  hline{26, 49, 51,54} = {2-9}{},
}
\textbf{Category}         & \textbf{Strategy}          & \textbf{Methodology}       & \textbf{Language}~\tnote{1}   & \textbf{Fix Level}~\tnote{2}  & \textbf{Repair Input}~\tnote{3}  & \textbf{Vulnerability Target}~\tnote{4}                                                & \textbf{Test Suites}~\tnote{5}  & \textbf{References  }                                                                                 \\
{Template\\Guided}   &                    &                  & C/C++      & S         & FC                 & GN                                                                 & \XSolidBrush         & \citep{weimer2006patches}                                                                     \\
                 &                    &                  & C/C++      & S         & S                  & Buffer Overflow                                                    & \XSolidBrush         & \citep{lin2007autopag, shaw2014automatically, gao2016bovinspector,huang2019using}             \\
                 &                    &                  & C/C++      & S         & P                  & Memory Error                                                       & \XSolidBrush         & \citep{novark2007exterminator, gao2015safe, hong2020saver,yan2016automated}                   \\
                 &                    &                  & C/C++      & S         & P                  & Integer Overflow                                                   & \XSolidBrush         & \citep{cheng2019automatic,muntean2019intrepair}                                               \\
                 &                    &                  & C/C++      & S         & S                  & Integer Overflow                                                   & \XSolidBrush         & \citep{coker2013program, huang2019using}                                                      \\
                 &                    & Vulnerability~   & Rust       & S         & P                  & Buffer Overflow                                                    & \XSolidBrush         & \citep{hua2021rupair}                                                                         \\
                 &                    & Property         & C/C++      & S         & FC                 & Error Handling Bugs                                                & \XSolidBrush         & \citep{tian2017automatically}                                                                 \\
                 &                    &                  & Solidity   & BT        & FC                 & Reentrancy, Acc Ctrl, Arithmetic, Uncheck LL Calls  & \XSolidBrush         & \citep{zhang2020smartshield, rodler2021evmpatch, ferreira2022elysium}                         \\
                 &                    &                  & Solidity   & S         & FC                 & Arithmetic, Reentrancy                               & \XSolidBrush         & \citep{nguyen2021sguard}                                                                      \\
                 & Template           &                  & Solidity   & S         & S, FC, P           & Reentrancy, Miss.Input.Vald., Lock.Ether, Unhandl.Except. & \XSolidBrush         & \citep{fang2023contractfix}                                                                   \\
                 & Construction       &                  & Java       & S         & S                  & SQL Injection                                                      & \XSolidBrush         & \citep{abadi2011automatically}                                                                \\
                 &                    &                  & Java       & S         & P                  & Null Pointer Deref.                                          & \XSolidBrush         & \citep{xu2019vfix}                                                                            \\
                 &                    &                  & /          & S         & S                  & Denial of Service                                                        &  \XSolidBrush         & \citep{li2022regexscalpel}                                                                            \\
                 &                    &                  & Java       & S         & P                  & Cryptographic Misuses                                              & \XSolidBrush         & \citep{ma2016cdrep, zhang2022example}                                                                           \\
                 &                    & Summaried~       & Java       & S         & P                  & GN                                                                 & \XSolidBrush         & \citep{ma2017vurle}                                     \\
                 &                    & From             & Java       & S         & S                  & GN                                                                 & \XSolidBrush         & \citep{kim2013automatic}                                                                      \\
                 &                    & Patches          & C/C++/Java & S         & P                  & Null Pointer Deref.                                          & \XSolidBrush         & \citep{xing8if}                                                                               \\
                 &                    &                  & C/C++      & B         & P                  & GN                                                                 & \XSolidBrush         & \citep{duck2020binary}                                                                        \\
 {Search\\Based}    &                    & Random           & C/C++      & S         & S                  & GN                                                                 & \CheckmarkBold       & \citep{weimer2009automatically,forrest2009genetic,le2012systematic,le2011genprog,tan2016anti} \\
                 &                    & Mutation         & C/C++      & S         & P                  & GN                                                                 & \CheckmarkBold       & \citep{zhang2022program}                                                                      \\
                 & Mutation~          &                  & Java       & S         & FL                 & Null Pointer Excep.                                             & \CheckmarkBold       & \citep{marginean2019sapfix}                                                                   \\
                 & Based              & Pattern~         & C/C++      & S         & P                  & Information Flow Leakage                                           & \XSolidBrush         & \citep{mesecan2021hypergi}                                                                    \\
                 &                    & Driven~          & Java       & S         & P                  & Null Pointer Excep.                                             & \XSolidBrush         & \citep{durieux2017dynamic}                                                                    \\
                 &                    & Mutation         & Solidity   & S         & P                  & GN                                                                 & \XSolidBrush         & \citep{tolmach2022property}                                                                   \\
                 &                    & Security~        & PHP        & S         & P                  & Injection Vulnerabilities                                                & \XSolidBrush         & \citep{shi2022backporting}                                                                    \\
                 &   Semantic~        & Patch            & C/C++      & S         & S                  & GN                                                                 & \XSolidBrush         & \citep{shariffdeen2021automated, yang2023enhancing}                                           \\
                 &   Code            & Transplantation     & C/C++      & S         & P                  & GN                                                                 & \CheckmarkBold       & \citep{shariffdeen2020automated}                                                              \\
                %  &                    & ~                & Java       & S         & P                  & GN                                                                 & \XSolidBrush         & \citep{nguyen2010recurring}                                                                   \\
                 &    Search          &             & PHP        & S         & S                  & Access Control Bugs                                                & \XSolidBrush         & \citep{son2013fix}                                                                            \\
                 &                   & Code~       & C/C++      & B         & FC                 & GN                                                                 & \XSolidBrush         & \citep{xu2020automatic}                                                                       \\
                %  &                    &                  & C/C++      & S         & S                  & GN                                                                 & \CheckmarkBold       & \citep{ke2015repairing}                                                                       \\
                 & & Similarity & Java & S & S & Null Pointer Deref. & \XSolidBrush &\citep{van2018static} \\
               &                    &                  & C/C++ & S         & S                  & Resource/Memory Leak, Null Pointer Deref.                       & \XSolidBrush         & \citep{van2018static}                                                                         \\
{Constraint\\Based}  &                    &                  & C/C++      & S         & FC                 & Memory Leak                                                        & \CheckmarkBold       & \citep{lee2018memfix}                                                                         \\
                 &                    & Static~ ~        & C/C++      & B         & FC                 & GN                                                                 & \CheckmarkBold       & \citep{duan2019automating}                                                                    \\
                 &                    & Analysis         & /          & S         & P                  & Null Pointer Deref., Data Leakage                            & \XSolidBrush         & \citep{liu2023program}                                                                        \\
                 &                    &                  & C/C++      & S         & P                  & Null Pointer Deref., Data Leakage                            & \XSolidBrush         & \citep{xing8if}                                                                               \\
                 & Constraint~ ~      &                  & /          & S         & S                  & Denial of Service                                                                & \CheckmarkBold       & \citep{chida2022repairing}                                                                    \\
                 & Extraction         &                  & C/C++      & S         & FC                 & GN                                                                 & \CheckmarkBold       & \citep{nguyen2013semfix, mechtaev2016angelix, shariffdeen2021concolic}                        \\
                 &                    & Symbolic         & C/C++      & S         & P                  & GN                                                                 & \CheckmarkBold       & \citep{gao2021beyond}                                                                         \\
                 &                    & Execution        & C/C++      & S         & S                  & GN                                                                 & \CheckmarkBold       & \citep{ke2015repairing, mechtaev2015directfix}                                                \\
                 &                    &                  & C/C++      & B         & FC                 & GN                                                                 & \XSolidBrush         & \citep{chen2017adaptive}                                                                      \\
                 &                    & Dynamic~         & C/C++      & S         & P                  & GN                                                                 & \CheckmarkBold       & \citep{gao2021beyond, gao2019crash}                                                                         \\
                 &                    & Analysis         & C/C++      & S         & P                  & GN                                                                 & \CheckmarkBold       & \citep{xuan2016nopol}                                                                         \\
                 &                    &          & C/C++      & B         & FC                 & GN                                                                 & \XSolidBrush         & \citep{xu2020automatic}                                                                       \\
                 &                    & Formal~           & C/C++      & S         & P                  & GN                                                                 & \CheckmarkBold       & \citep{zhang2022program}                                                                      \\
                 &                    &  Methods                &  /     & /         & /                 & Trigger Action Programming                                              & \CheckmarkBold       & \citep{yu2024tapfixer}                                                                       \\
{Learning\\Based}                 & Training           & Deep Learning         & C/C++      & S         & FC                 & GN                                                                 & \XSolidBrush         & \citep{chen2022neural, harer2018learning}                                                     \\
                 &                    &                  & Java       & S         & S                  & GN                                                                 & \XSolidBrush         & \citep{chi2022seqtrans}                                                                       \\
         & Fine-tuning        & Adaption         & C/C++      & S         & S                  & GN                                                                 & \XSolidBrush         & \citep{huang2022repairing}                                                                    \\
            &             & Interaction      & C/C++      & S         & FC                 & GN                                                                 & \XSolidBrush         & \citep{islam2024code}                                                                         \\
                 &   Prompt                        & Zero-shot        & C/C++/Java & S         & FC                 & GN                                                                 & \XSolidBrush         & \citep{pearce2023examining, wu2023effective, wu2023exploring, ahmad2023fixing, wang2024navrepair, liu2024prompt, liu2024exploring}                \\
                 &   Engineering                 & Few-shot         & C/C++      & S         & FC                 & GN                                                                 & \XSolidBrush         & \citep{fu2023chatgpt, fu2024vision}                                                                         \\
                 &                    & Chain-of-thought & C/C++/Java/Solidity      & S         & FC                 & GN                                                                 & \XSolidBrush         & \citep{nong2024chain, khan2024code,wang2024contracttinker, kulsum2024case}                                                              \\           
                 &  LLM                &   Multi-LLM               & C/C++/Java      & S         & FC                 & GN                                                                 & \XSolidBrush         & \citep{zhou2024out}                                               \\                          
                 &  Integration                  &    LLM-External Tool              & C/C++/Java      & S         & FC                 & GN                                                                 & \CheckmarkBold         & \citep{tihanyi2023new, kulsum2024case}                           \\                                              
                \end{tblr}
}
\begin{tablenotes}
\small{
  \item[1] \textbf{Language}: Programming language that SPG method targets at, ``/" indicates no specific language is targeted or not a programming language.
  \item[2] \textbf{Fix Level}: The level at which security patches are generated: "S" for source code level, "B" for binary, and "BT" for bytecode.
  \item[3] \textbf{Repair Input}: Granularity of Input considered for SPG, from statements (S), functions (FC), and files (FL), to projects (P).
  \item[4] \textbf{Vulnerability Target}: Classifies methods by the vulnerabilities they address. ``GN'' means not target specific vulnerabilities. 
  \item[5] \textbf{Test}: Whether test suites are required for SPG. ``\CheckmarkBold''  indicates test suites are needed, ``\XSolidBrush'' means they are not.}
 \end{tablenotes}
\end{threeparttable}
\vspace{-1em}
\end{table*}

\begin{figure*}
  \centering
  \includegraphics[width=1.03\textwidth]{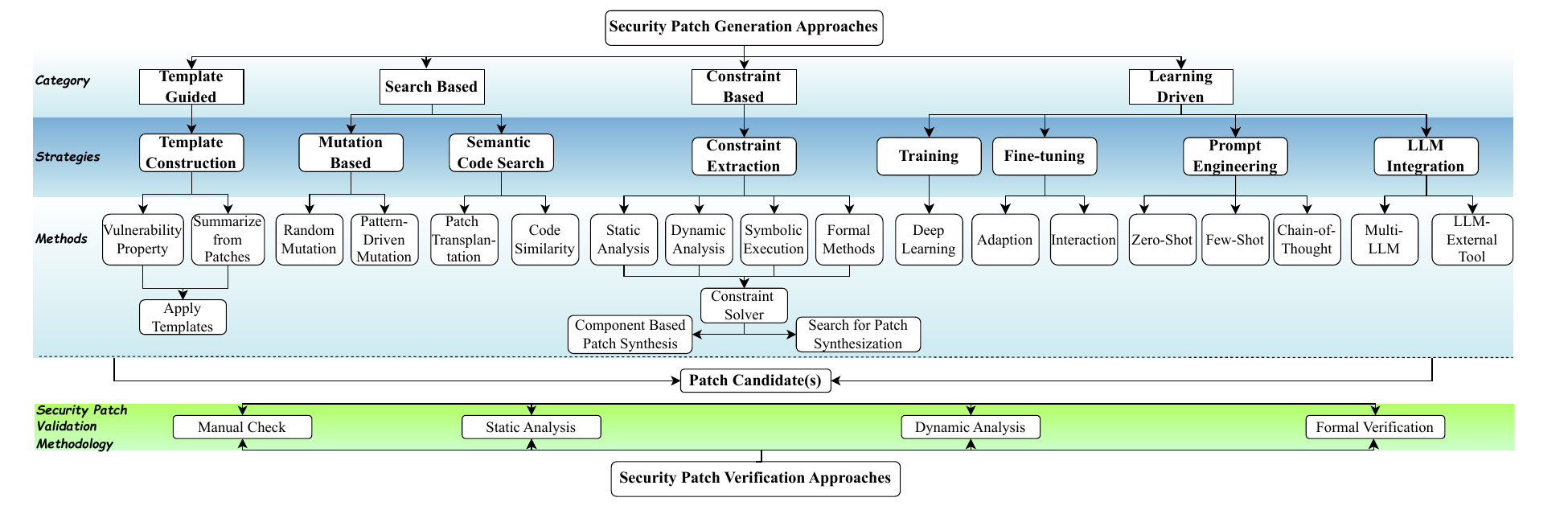}
  \caption{\textbf{Taxonomy} of Automated Vulnerability Repair approaches. 
  % \textcolor{blue}
  Boxes with \textit{blue} background present the taxonomy of security patch generation methods. Boxes with \textit{green} background show the taxonomy of security patch validation approaches.
 }
  \label{fig:taxonomy}
  \vspace{-0.3cm}
\end{figure*}

% \vspace{-1em}
\section{Taxonomy of Automatic Vulnerability Repair Approaches}
\label{sec:taxonomy_apr}

\newtext{\subject{Selection Methodology.} We conducted a systematic literature search focused on AVR using four search queries: ``security patch generation'', ``vulnerability repair'', ``program repair + vulnerability'' and ``vulnerability patching''. The search was performed on\url{cspapers.org} automatically, limiting results to the domains of Computer Security, Software Engineering, and Programming Languages. Initial results yielded 4,883 unique papers published between 2018 and 2024.9. To increase relevance, we applied filtering criteria requiring papers to have a relevance score above 0.5 and contain at least one of the following keywords in their abstracts: ``security'', ``vulnerability'', ``repair'', ``patch'', or ``fix''. This filtering process resulted in 267 unique papers. These papers underwent detailed review by two authors, spending about 45 minutes per paper to determine if the paper targets AVR, with weekly consensus meetings to resolve disagreements. Through this manual review process, we identified 23 core papers for AVR. Finally, we applied both forward and backward snowballing like~\citep{lefeuvre2024sok, ladisa2023sok} to examine referenced and citing papers, identifying additional 56 AVR papers, which cross 2006 to 2024.9, the process terminated when no new methodologies, strategies, programming languages on AVR were found in two consecutive iterations. The final collection comprises \papernumber papers, with \Anumber ranked A/A* on CORE2023~\cite{core}, a computing research venue ranking system where A/A* represents the top 20\% venues. %in terms of impact and prestige. 
}

\newtext{\subject{Analysis Methodology.} To analyze each paper, we employed a systematic approach to analyze each paper, examining three key dimensions: (1) The core components and methodologies used in AVR; (2) The effectiveness and target vulnerabilities types; (3) The strengths and limitations of each approach. We analyze each paper using a standardized rubric evaluating technical architecture, vulnerability types addressed, repair strategies with their theoretical foundations, experimental validation, and limitations. Uncertainty and inconsistencies were resolved through group discussion. This analysis revealed fundamental distinctions between learning-based and non-learning approaches. Within non-learning approaches, we identified three distinct categories: template-guided approaches using predefined templates, search-based approaches exploring security patch spaces or vulnerable states, and constraint-based approaches that formalize repairs through constraint extraction and solving. Further analysis within each category revealed distinct strategies, leading to our complete hierarchical taxonomy.
}

Figure~\ref{fig:taxonomy} presents a hierarchical taxonomy, categorizing repair approaches into SPG and SPV (assuming successful vulnerability localization).
SPG approaches are further classified into three levels: (i) broad categories (Template-Guided, Search-Based, Constraint-Based, and Learning-Driven), (ii) specific strategies within each category, and (iii) core methodologies. A detailed analysis of these approaches is provided in Section~\ref{sec:patch_generation}. \newtext{SPV, classified into manual check, static analysis, dynamic analysis, and formal verification, shows a strong presence of manual in AVR practices (\manualcnt out of \papernumber AVR papers), either as a standalone approach or in combination with other methods. While VL is a critical prerequisite step for AVR, AVR papers often stress their core contribution to SPG, with 15 papers even assume known vulnerability locations.}

\subject{Properties of Security Patch Generation approaches.} From Table~\ref{tab:gen_taxonomy}, we observe different features under different categories. \newtext{For \textbf{Language}, 66 out of \papernumber papers focus on C/C++/Java, others focus on other languages (e.g., Rust). Notably, one paper~\cite{yu2024tapfixer} focus not on programming language but on natural language. This language distribution aligns with real-world trends: C/C++ dominate memory-safety issues~\cite{sakharkar2023systematic, mm}, while Rust sees limited AVR coverage due to its newness.} For \textbf{Fix Level}, only \bincnt provide AVR at the binary level, despite the importance of binary-level defense in deployed software systems~\cite{shariffdeen2022software}.
Notably, 3 out of 6 non-learning-based studies on Solidity focus on bytecode-level instead of source-code level due to source code availability (0.3\%)~\cite{ferreira2022elysium}. For \textbf{Repair Input}, current learning-based methods typically operate at statement-level or function-level repair input. This contrasts with other approaches, which offer support across all four input levels for repair, depending on the specifics of their implementation.  Additionally, in the process of generating security patches, both learning-based and template-guided methods do not require the involvement of test cases or exploits. In contrast, other approaches may necessitate test cases/exploits, depending on their specific implementation details. For \textbf{Vulnerability Target}, 85.71\% of the template-guided methods target one or more specific vulnerability types.

Various methods exhibit distinct strengths and weaknesses across different properties. For example, source-code level fixes are more straightforward but binary-level fixes save compilation time~\citep{duck2020binary}. Function/statement-level repair input is faster than program-level but lacks global program context. Among \nongncppcnt papers targeting specific vulnerabilities in C/C++, 62.50\% focus on memory-related vulnerabilities such as memory leak and buffer overflow, which are crucial because exploiting them could give attackers access to the whole system~\citep{erickson2008hacking}. However, few works focus on logic vulnerabilities, which are equally important, if not more so (e.g., DAO attack~\cite{dao} results in \$8.5 million ETH loss). Detailed pros and cons of each method will be discussed in Section~\ref{sec:patch_generation}.

% Technique-wise:

% 1. Static

% - Rule-based

%     ----- Heuristic
     
%     ----- Constraint-based 
     
% - Program Analysis 

% 2. Dynamic

% - ML

% - DNN

% - LLM

\section{Security Patch Generation Approaches}
\label{sec:patch_generation}

In this section, we present typical SPG approaches: \textit{Template-Guided}, \textit{Search-Based}, \textit{Constraint-Based}, and \textit{Learning-Driven} methods. 
We conclude each subsection by highlighting the \textit{Takeaways} and \textit{Open Problems} in this section.

\subsection{Template-Guided }

Template-Guided approaches for SPG rely on predefined repair templates or safety properties derived from human expertise and vulnerability characteristics. Their efficacy depends on the accuracy and coverage of the identified templates or properties, which determine the scope and precision of vulnerabilities that can be fixed. This section explores how various research efforts define and utilize these templates or properties to remediate security vulnerabilities.

\subject{A. Template Construction.}
There are two main directions for constructing the template: 

\subsubject{\underline{A1. Templates based on Vulnerability Property.}} 
Different types of vulnerabilities have unique root causes and characteristics. 
Customizing security properties to address these facilitates the development of efficient repair templates.
This approach ensures that each vulnerability is treated with a strategy that aligned to its nature.
Initially, Weimer~\cite{weimer2006patches} proposed using an abstract behavioral model to generate security patches. Later, AutoPag~\cite{lin2007autopag} developed templates specifically for out-of-bound vulnerabilities, but it was only applicable when the vulnerability and the fix location were in the same function. 
Shaw et al.~\cite{shaw2014automatically} focused on replacing unsafe APIs with safer alternatives to fix buffer overflows, but this template is effective only when the vulnerabilities are directly introduced by these unsafe APIs. Furthermore, BovInspector~\cite{gao2016bovinspector}, Senx~\cite{huang2019using}, and INTREPAIR~\cite{muntean2019intrepair} utilized safety properties of corresponding vulnerabilities related to specific vulnerabilities to derive patches, including buffer overflow, bad casts, integer overflows, and dangling pointers. 
Besides repairing statically, Exterminator~\cite{novark2007exterminator} fixes memory bugs at runtime without requiring source code changes. \newtext{Besides memory vulnerabilities, RegexScalpel~\cite{li2022regexscalpel} applies predefined repair templates (e.g., modifying quantifiers) to fix Regular Expression Denial of Service (ReDoS) vulnerabilities (e.g., nested quantifiers).}

Templates derived from vulnerability properties can be represented or mined from various program analysis graphs (e.g., control flow graph, program dependency graph, abstract syntax tree (AST))~\cite{allen1970control, ferrante1987program}.  These approaches primarily involve traversing the graph and modifying elements in fixed locations within the graph to achieve security patch generation.
The basic idea has been used to fix common vulnerabilities, including memory management bugs (e.g., memory leak, double free, use after free ~\cite{gao2015safe, yan2016automated}), error handling bugs~\cite{tian2017automatically}, null pointer dereferences~\cite{xu2019vfix}, and integer overflows~\cite{coker2013program, cheng2019automatic}. 
Furthermore, SAVER~\cite{hong2020saver} formulated the vulnerability fix the problem as a graph labelling problem to implement fixes.

Although the above researches have defined patterns to fix buffer overflow in C/C++/Java, the approaches cannot be directly used for fixing Rust buffer overflow vulnerabilities due to its unique language features like ownerships and lifetimes~\cite{ownership}.
Also, Rust vulnerabilities arise from interactions between its safe and unsafe sub-languages~\cite{vadayatharbiter, jung2019stacked,qin2020understanding}.
Rupair~\cite{hua2021rupair} defines the patterns to rectify the vulnerabilities by semantics-preserving program transformations. 
Similarly, the approaches are also not applicable on Solidity \newtext{due to their domain-specific vulnerabilities, such as reentrancy attacks, gas limit issues, which are not common in other languages in traditional software. 
SGUARD~\cite{nguyen2021sguard} and CONTRACTFIX~\cite{fang2023contractfix} apply fixing templates to the source code to ensure the smart contract is free from those vulnerabilities, while EVMPatch~\cite{rodler2021evmpatch}, Smartshield~\cite{zhang2020smartshield}, and Elysium~\cite{ferreira2022elysium} operates on bytecode level. However, they follow similar strategies, for example, SGUARD~\cite{nguyen2021sguard} add \texttt{nonReentrant} modifier for reentrancy vulnerabilities, EVMPatch~\cite{rodler2021evmpatch} added checks to verify the caller's address or permission on bytecode level. }

\subsubject{\underline{A2. Templates Summarized from Patches.}} 
Different from templates guided by vulnerability properties, templates summarized from existing patches focus on generalizing successful strategies from real-world fixes.
PAR~\cite{kim2013automatic}, Vurle~\cite{ma2017vurle}, Seader~\cite{zhang2022example}, and CONCH~\cite{xing8if} summarized fix templates from patches for SPG. Ma et al.~\cite{ma2016cdrep} summarized cryptographic misuses and corresponding fix patterns in Android. E9PATCH~\cite{duck2020binary} defines four tactics for binary rewriting, proving effective in repairing vulnerabilities in binaries.

\subject{B. Template Application.} Once constructed, the template is applied for patch generation. Note that not all templates are in source code format. When a template is not in source code, it must be mapped to concrete expressions (e.g., Senx~\cite{huang2019using}) to facilitate the generation of a patch in source code.

\takeaways{A template, whether enforcing security properties or serving directly as a repair template, often incorporates heuristic rules to address specific vulnerabilities or security property violations. This approach is effective to fix vulnerabilities that require minimal changes, have limited context dependency, and have clear root causes, such as switching to safer APIs or changing array sizes.
While template-based methods lack flexibility, they are still a great starting point when applied to new contexts, such as binary-level repair~\cite{duck2020binary}. 
Furthermore, templates can work with other repair methods throughout the AVR process, such as by extracting specific types of constraints~\cite{gao2021beyond} to enhance SPG.
}

\openquestion{  
\begin{itemize}[leftmargin=*]
 \item \textbf{Template Mining:} 
Most current template-guided AVR methods rely on hard templates (pre-defined and unchanging repair strategies)\newtext{(e.g.,~\citep{kim2013automatic, hong2020saver, hua2021rupair,zhang2020smartshield,rodler2021evmpatch,nguyen2021sguard,fang2023contractfix}). While same type vulnerabilities often follow shared patterns, their specific manifestations can vary due to evolving contexts or exploitation techniques, necessitating flexible soft templates~\cite{kim2013automatic, ma2017vurle, zhang2020smartshield}}. While existing research on patch generation explores using models to create repair templates, vulnerabilities often involve complex dependencies that require more flexible approaches. Thus, a soft template mining method is urgently needed.
 \item \textbf{Security Property Inference:}
 Security properties~\cite{huang2019using} can be viewed as templates to enforce stringent security conditions. Unlike typical fix templates, they guide other methods (e.g., solvers) to generate patches that meet complex requirements, such as sequence dependencies. Inferring these properties from various sources is essential for ensuring robust SPG. %which often involves formal methods.
 \end{itemize}
 }

\subsection{Search Based}
\label{sub:search}

% \fayrew{
Heuristically search-based SPG is guided by defining heuristic rules to search and generate patches~\cite{heuristic_search}.
Two primary approaches within this category are \textit{mutation-based} and \textit{semantic code search}. Both establish a search space and select strategies accordingly. Mutation-based approach modify specific code sections for SPG. Conversely, the \textit{semantic code search} seeks to identify code segments that are semantically similar  but non-vulnerable code segments as potential solutions.
% }

\subject{A. Mutation-Based Approach.} This approach focus on modifying operations within a designated statement at a specified patch location, primarily by transforming the program’s AST. Mutations at the chosen repair site, involve alterations to operators, variables, APIs, and operation types within the AST. These mutations fall into two categories: \textit{random mutation}, where changes are applied randomly, and \textit{pattern-driven mutation}, guided by predefined patterns or rules.

\subsubject{\underline{A1. Random Mutation.}} Random mutation introduces a level of randomness during mutation, primarily using genetic programming and fuzzing.
Genetic programming fundamentally operates through a cycle of $mutation \leftrightarrows test$. This cycle continues until the mutated program successfully passes all test suites.
GenProg~\cite{forrest2009genetic, weimer2009automatically, le2011genprog, le2012systematic} implements this concept, guided by the heuristic that ``a program containing an error in one context likely implements correct behaviour elsewhere''~\cite{engler2001bugs}. This involves modifying the AST and recombining existing code snippets through mutation until all test suites are passed. This technique has demonstrated effectiveness in repairing vulnerabilities such as buffer overflow and format string issues~\cite{forrest2009genetic, weimer2009automatically, le2011genprog, le2012systematic}. 
However, the vast search space often leads to low-quality patches. To improve this, Tan et al.~\cite{tan2016anti} suggested norrowing the search space by capturing disallowed modifications (i.e., modifications considered inappropriate or counterproductive for fixing bugs). However, obtaining ``complete'' anti-patterns is challenging. Anti-patterns are captured based on their frequency in bad patches and their rarity in correct ones. Consequently, patterns that are less prevalent in bad patches might be omitted. 

Apart from genetic programming, fuzzing~\cite{fuzz} is another technique for random mutation. VulnFix~\cite{zhang2022program} employs snapshot fuzzing to mutate program states directly while inferring patch invariants. These invariants then guide the constraints on the patches, offering a more targeted approach for AVR.

\subsubject{\underline{A2. Pattern Driven Mutation.}}  
Pattern-driven mutation employs generalized patterns for modifications, such as adding null pointer checks, modifying boolean expressions, or altering arithmetic operations. The mutation patterns address many issues but may lack proven effectiveness for specific cases. In contrast, template-guided repair uses templates that have been proven effective in other fixes. 
Sapfix~\cite{marginean2019sapfix} employs two mutation strategies involving statement insertion, specifically for \texttt{return null} or \texttt{null check} operations. Given that information leakage is closely tied to control flow~\cite{he2020ct}, HyperGI~\cite{mesecan2021hypergi} uses standard operators along with two novel ones like \texttt{|if|}  and \texttt{|for|} to perform control flow-dependent mutations.
However, these fixes are performed statically,  Durieux et al.~\cite{durieux2017dynamic} explore the search space of possible patches for null pointer exceptions with metaprogramming~\cite{meta_programming}, which analyzes the program's real-time behavior and selects a repair pattern to mutate based on the current execution context.
Beyond the realm of C/C++ and Java vulnerability repair, there is a growing interest in fixing vulnerabilities in Solidity, particularly for DeFi protocols. DeFinery~\cite{tolmach2022property} adopts three employs three operator mutation rules for this, notably without needing access to test cases or historical transaction data.

\subject{B. Semantic Code Search.}
\label{sub:scs}
When code exhibits similar vulnerabilities in comparable contexts or parallel functions, analogous patching strategies are often necessary. \newtext{Since 20\%-50\% of large software systems consist of cloned code (i.e., code segments that are duplicated or highly similar across different parts of a software project)~\citep{baker1995finding, ducasse1999language}, and patches are generally developed for the most recent version~\cite{patch-version}, }
% This is especially prevalent in Open Source Software, where code cloning is common~\cite{lozano2008assessing}, 
earlier versions often have vulnerabilities that are contextually similar to these updates. Furthermore, software vulnerabilities frequently recur across different applications~\cite{pham2010detection}. 
Consequently, by analyzing the characteristics of existing patches for known vulnerabilities or studying benign code that lacks these vulnerabilities, it becomes possible to repair similar unpatched vulnerabilities. Inspired by patch semantics, this approach has been applied in \textit{Security Patch Transplantation} and \textit{Security Patch Generation based on Similarity}.

\subsubject{\underline{B1. Security Patch Transplantation.}}
Patch backporting is a key aspect of security patch transplantation~\cite{patch_backport}, adapting the latest official patches for known vulnerabilities to older software versions, ensuring both compatibility~\cite{padioleau2008documenting, thung2016recommending} and continued security~\cite{linux_backport}.
Tools like SKYPORT~\cite{shi2022backporting}, FIXMORPH~\cite{shariffdeen2021automated}, and TSBPORT~\cite{yang2023enhancing} are developed for this purpose. SKYPORT targets web applications, whereas TSBPORT and FIXMORPH target Linux kernel vulnerabilities. 
These tools assume that official patches are semantically consistent across versions, but semantic differences can challenge automated patching, even when the exact fix location is known~\cite{yang2023enhancing}.
Another approach within security patch transplantation is to spread fixes across various implementations of the same protocol or functionality. For instance, PatchWeave~\cite{shariffdeen2020automated} uses concolic execution to find and apply matching symbolic expressions in different software, enabling the transfer of fixes to similar issues in diverse environments.

\subsubject{\underline{B2. Security Patch Generation based on similarity.}}
Unlike patch transplantation, vulnerabilities often reappear across various software systems and functionalities~\cite{le2016history}. Thus, historical fixes serve as valuable references for addressing similar vulnerabilities. The key idea is to identify and apply similar patches to fix analogous issues. 
FixMeUp~\cite{son2013fix} targets faulty access-control logic in web applications and inserts similar code where needed.
% Other methods, like Ke et al.~\cite{ke2015repairing} and 
Van et al.~\cite{van2018static}, match code fragments to ensure functional properties or explore similar code within the same program that meets necessary conditions. However, these repair methods require system interruption during patching.
Later, Xu et al.~\cite{xu2020automatic} determine the similarity between the code to be fixed and the buggy code by comparing the semantics of the official patches with the vulnerable functions using program analysis and weakest precondition reasoning.

\takeaways{Mutation-based SPG uses random mutations that may not specifically tailored vulnerabilities but effective with clear patterns. Semantic code search uses related patches as references, but relies on robust matching. Similarly, similarity-based SPG explores large search spaces~\cite{wen2018context} by detecting code similarities, requiring sound verification (e.g., comprehensive test suites). }

\openquestion{\begin{itemize}[leftmargin=*]
    \item \textbf{\newtext{Search Space Optimization.}} 
    % In search-based vulnerability repair, a spellbinding research opportunity lies in the exploration of the ``optimal search space''. 
    \newtext{How can we automatically optimize the search space and adapt search strategies based on different scenarios (e.g., different types of vulnerabilities, different settings, i.e., across projects or within a project)~\cite{forrest2009genetic, weimer2009automatically, le2011genprog, le2012systematic, tan2016anti}? }
    \item \textbf{Sparse Vulnerability Patterns.} \newtext{How can we enhance the search methods for rare vulnerabilities~\cite{nguyen2010recurring}, leveraging the unique properties of these vulnerabilities to guide program synthesis? } Additionally, how can we leverage the unique properties of these rare vulnerabilities to guide the program synthesis effectively?
    \item \textbf{Complex Logic and Structural Changes in Backporting.} \newtext{How can we effectively backport patches involving complex logic or structural changes,  considering the need to understand intricate program semantics, handle significant alterations to algorithms or data structures, and dependencies or contexts that may not exist in the target version~\cite{shariffdeen2021automated,shi2022backporting,yang2023enhancing, shariffdeen2020automated}? } 
\end{itemize}}

\subsection{Constraint Based }
% Security Patch Generation}

Program constraints are limitations or conditions imposed on program behavior, inputs, outputs, or data structures~\cite{marriott1998programming}. 
In constraint-based SPG, the key lies in formalizing the constructed or extracted constraint that the synthesized patch is required to satisfy. The methods focus on formal constraint generation rather than constraint solving~\cite{goues2019automated}. 
% In constraint-fundamental aspect of constraint-based patch generation approaches involves formalizing the constructed or extracted constraint that the synthesized patch is required to satisfy. 
% In these methods, the key does not lie in constraint solving but in formal constraint generation\cite{goues2019automated}. 
Constraint-based approaches could be summarized into two main steps: (i) Extract repair constraints. (ii) Employ constraint solvers or search for statements that fulfill the constraints. Our discussion is organized around various methods of the two steps. 

\subject{A. Constraint Extraction.} Constraint extraction involves analyzing the affected code to identify the specific requirements that any potential patch must adhere to. Constraint extraction methods can be classified into four approaches.
% \ying{Add 1/2 line description here, todo}

% \yuan{Why don't we cover all the 3 step?}\ying{rigrously, it should be 2 steps, fixed.}

\subsubject{\underline{A1. Static Analysis Driven Constraint Extraction.}} 
Utilizing static analysis could identify constraints related to variables' values, their interrelationships, and the function calls sequences. These identified elements can then be utilized to generate further constraints. MemFix~\cite{lee2018memfix} uses typestate analysis~\cite{fink2008effective} to track memory object states across branches, and generate patches for each state. It frames the problem as an NP-complete exact cover issue~\cite{exact_cover}, focusing on patching memory vulnerabilities (e.g., double free). However, it is limited to modifying deallocation statements, excluding conditionals.
CONCH~\cite{xing8if} employs state propagation, utilizing CFG to encompass the full call chain for constraint extraction. This strategy aims to craft patches for \newtext{Null Pointer Deferences (NPDs)}. Chida et al.~\cite{chida2022repairing} introduced a method by analyzing templates to maintain condition consistency with examples to generate constraints for repairing regex denial-of-service vulnerabilities. 
Symlog~\cite{liu2023program} employs Datalog-based~\cite{datalog} static analysis to generate structural constraints directly, effective across multiple languages due to Datalog's declarative nature. However, the above methods only work when the application is open-source and focuses on source code. 
OSSPATCHER~\cite{duan2019automating} assumes closed-source applications with available patches from open-source code.  
It employs a variability-aware AST to derive constraints, such as macro definitions and conditional compilations, and adapts these patches into binary form, aligning them with extracted constraints.

\subsubject{\underline{A2. Symbolic Execution Driven Constraints Extraction.}} 
Symbolic execution~\cite{king1976symbolic} executes the program with symbolic values, enabling the computation of constraints.
Depending on the different objects of focus, different types of constraints can be computed through symbolic execution via tools such as KLEE~\cite{cadar2008klee}. 
The most representative method is to adopt symbolic execution to capture program semantic constraints that could pass all test cases~\citep{nguyen2013semfix, mechtaev2015directfix, mechtaev2016angelix}.
Further, the constraint for memory writes and function calls~\cite{chen2017adaptive}, the weakest precondition, the extension of crash-free constraints~\cite{gao2021beyond} could also be computed by symbolic execution.
To achieve better path coverage, concolic execution~\cite{godefroid2005dart}, which employs concrete input to drive symbolic execution, is also utilized to traverse the path driven by path constraints~\cite{shariffdeen2021concolic}.

\subsubject{\underline{A3. Dynamic Analysis Driven Constraints Extraction.}}
Dynamic analysis extracts runtime constraints by observing program behavior and potential vulnerabilities during execution. 
NOPOL~\cite{xuan2016nopol} collects runtime data during test case execution to analyze the program behavior, identifying relevant variables and expressions, and encoded into a Satisfiability Modulo Theory (SMT) problem~\cite{barrett2018satisfiability}.
However, this method relies heavily on test suites quality, \newtext{and most of the time, vulnerabilities may only have one exploit}. To address this,  ExtractFix~\cite{gao2021beyond} generates patches that generalize beyond a single exploit trace. Specifically, it adopts sanitizers~\citep{serebryany2012addresssanitizer, undifiedbehavior} to convert vulnerabilities into normal program crashes, allowing the extraction of precise condition constraints at the crash point.

\subsubject{\underline{A4. Formal Methods for Constraints Extraction.}} Formal methods use mathematical techniques and theories to verify and prove software correctness~\cite{butler2001formal}. These methods involve constructing mathematical models that describe the behavior and properties of a program, allowing for the definition of precise constraints that the software should satisfy. 
VULMET~\cite{xu2020automatic} employs the weakest precondition reasoning to transform known patches into constraints. VulnFix~\cite{zhang2022program} derives repair constraints via inductive inference~\cite{solomonoff1964formal}. \newtext{VulnFix begins by collecting program state snapshots at vulnerability location during benign and vulnerable executions, then use Daikon~\cite{daikon} to infer initial invariants that distinguish benign states from vulnerable ones and CVC5~\cite{barbosa2022cvc5} verifies and refines these invariants to enhance the accuracy}. Unlike approaches dependent on symbolic and concolic execution, these methods scale better to larger programs. \newtext{Additionally, vulnerabilities not only exist in the code, but also stem from higher-level abstractions, like Trigger Action Programming (e.g., turn off lights when idle may conflict with a rule to keep them on), where logical inconsistencies or design flaws can introduce security risks. TAPFixer~\cite{yu2024tapfixer} detects rule conflicts via model checking~\cite{clarke1997model} and resolves them using Negated-Property Reasoning, which analyzes violated properties to generate fixes such as adding delays, modifying conditions.} 

\subject{B. Patch Generation Based on Constraints.}
After extracting repair constraints, two typical methods for generating patches emerge. The first involves synthesizing a patch by finding a solution that ensures correct behavior and avoids the vulnerable state, typically using a component-based program synthesis approach. The second method involves searching the codebase for code that meets these constraints.

\subsubject{\underline{B1. Component-based Patch Synthesis}} involves selecting and combining predefined components (e.g., variables, operations) to automatically generate patches that satisfy expected behavior, typically using SMT solvers. 
The choice of SMT solvers depends on the constraint features and problem objectives. General-SMT solvers, like Z3~\cite{de2008z3}, are widely used in current research and applications, as evidenced by their use in a variety of works~\citep{lee2018memfix, duan2019automating, liu2023program, nguyen2013semfix, xuan2016nopol, mechtaev2016angelix, chen2017adaptive, gao2021beyond, shariffdeen2021concolic}, in which~\cite{duan2019automating} implement source-to-binary matching. 
For example, in \texttt{strcpy} function \cite{strcpy}, a buffer overflow occurs if the \texttt{source} parameter is larger than \texttt{destination}. To prevent this, a constraint ensures \texttt{length(source) <= length(destination)}. 

Besides checking the satisfiability of logical formulas over one or more theories, SAT~\cite{sat} solvers handle Boolean logic satisfiability checks. Gopinath et al.\cite{gopinath2011specification} translate the constraint \texttt{\small {$pre-state \wedge code-constraints \wedge post-condition$}} into boolean logic and solved by SAT solver.

However, when multiple patches meet the constraints using first-order logic, a second-order solver~\cite{mechtaev2018symbolic} enhances patch quality. For instance, EXTRACTFIX~\cite{gao2021beyond} employs pMaxSMT (partial MaxSAT) to handle both hard and soft constraints, while DirectFix~\cite{mechtaev2015directfix} transforms the patch generation task into a pMaxSAT problem, using patch complexity as a soft constraint. This approach helps produce simpler patches.

\subsubject{\underline{B2. Search Codebase For Patch Synthesization.}} Unlike \textit{component-based patch synthesis}, this approach searches codebase for existing code that satifies extracted constraints, bypassing the need to reassemble code components within the project~\cite{chida2022repairing}. 
More details are elaborated in Section~\ref{sub:scs}.

\takeaways{\newtext{
Constraint-based SPG excels in capturing precise vulnerability requirements and program-wide properties, ensuring fixes satisfy specified invariants or constraints. Unlike heuristic methods, this approach is particularly effective for vulnerabilities requiring rigorous analysis of safety properties.}}

\openquestion{\begin{itemize}[leftmargin=*]
    \item \textbf{Vulnerability-Specific Constraints.} High-quality constraints are the key to constraint-based SPG (e.g., ~\cite{chida2022repairing, liu2023program, duan2019automating}), \newtext{because it defines the properties that must hold true or false for repaired code. An open problem is how to accurately infer and differentiate between ``benign'' constraints and those specific to vulnerabilities.}
    \item \textbf{Security Specification Generation.} Constraints can be considered a concretization of the specification, further refined to guide the implementation and operation of the system.
    However, specifications are often incomplete or underspecified~\cite{specification-problem}, leading to ambiguity in the constraints to be enforced. Thus, an open challenge is to develop methods for inferring detailed specifications from existing code, textual information, etc. that can then be used to guide the constraint generation process.
\end{itemize}}

\subsection{Learning Driven }
\label{subsec:learning}

Learning-driven SPG aims at utilizing learning methods to perform end-to-end security patch generation. Although in general bug repair, learning methods have been used to rank patches~\citep{long2016automatic, saha2017elixir, white2019sorting, li2020dlfix}, extract templates for patch generation~\citep{rolim2017learning, koyuncu2020fixminer}, and end to end patch generation~\citep{gupta2017deepfix, vasic2019neural, chen2019sequencer, duan2019automating, tufano2019empirical, chakraborty2020codit, lutellier2020coconut, li2022dear}, their application in AVR remains primarily limited to SPG. 

Learning-driven SPG methods encompass \textit{Training}, \textit{Fine-tuning}, \textit{Prompt Engineering}, and \newtext{\textit{LLM Integration}, with the latter three relying on large (code) language models}. These models include encoder-decoder models that transform an input sequence $x=(x_{0}, x_{1},..., x_{T})$ into a corresponding output sequence; decoder-only models for next token prediction; and infilling models that generate contextual missing segments.

\subject{A. Training.} Training involves building a model from scratch. In SPG, deep learning is particularly favoured for this purpose.

\subsubject{\underline{\textit{A1. Deep Learning.}}}  In deep learning, Neural Machine Translation (NMT) becomes fundamental for SPG. It utilizes encoder-decoder architectures, often using RNNs or transformers, learn mappings between source and target code sequences. 
For instance, Seqtrans~\cite{chi2022seqtrans} and VRepair~\cite{chen2022neural} utilize NMT to learn rules from historical fixes
(including vulnerability fixes and bug fixes) 
and apply them to future edits. Additionally, Harer et al.~\cite{harer2018learning} tackle vulnerability repair using adversarial learning with Generative Adversarial Networks~\cite{goodfellow2020generative}. They utilize an NMT model as the generator within the GAN framework, enabling training without paired examples of code.

\subject{B. Fine-tuning.} Fine-tuning involves adjusting a pre-trained model to suit a specific task or domain. Since these models are not immediately suitable for AVR, significant customization and re-training are necessary to make them effective.

\subsubject{\underline{\textit{B1. Adaption.}}}  
One method for fine-tuning is adaption, which focuses on adapting knowledge from one domain to another.  Given the limit dataset for vulnerability fixes, the similarity between vulnerability fixes and bug fixing prompts the use of bug-fix datasets for pre-training~\cite{chi2022seqtrans, chen2022neural, zhang2023pre, zhou2024out}. 
However, self-trained models on bug-fixing datasets often fail to capture rich programming features~\cite{jiang2021cure}. Some pre-trained models, trained on large codebases, can be fine-tuned to effectively address vulnerabilities. For example, VULRepair~\cite{huang2022repairing} fine-tuned pre-trained Programming Language models for SPG.

\subsubject{\underline{\textit{B2. Interaction.}}} Fine-tuning can also involve learning through environmental interaction, i.e., reinforcement learning~\cite{kaelbling1996reinforcement}.
To ensure syntactic and semantic consistency, Islam et al.~\cite{islam2024code} used CodeBLEU~\cite{ren2020codebleu} score and BERTScore~\cite{zhang2019bertscore} as rewards, guiding fine-tuning with Proximal policy optimization~\cite{schulman2017proximal} to fine-tune CodeGen2 model~\cite{nijkamp2023codegen2} for SPG.

\subject{C. Prompt Engineering.} Prompt engineering guides LLMs to produce optimal SPG outputs through designed inputs, encompassing \textit{Zero-shot}, \textit{Few-shot}, and \textit{Chain-of-thought prompting}.

\subsubject{\underline{B1. Zero-shot prompting.}} Zero-shot prompting involves providing prompts without labeled data. A line of research~\cite{pearce2023examining, wu2023effective, wu2023exploring, ahmad2023fixing, wang2024navrepair, liu2024prompt, liu2024exploring} explores the capability of zero-shot AVR with different LLMs, including Codex~\cite{codex}, CodeT5~\cite{wang2021codet5}, GPT-4~\cite{gpt_4}, CodeGen2~\cite{nijkamp2023codegen2} etc. Among these, Pearce et al.~\cite{pearce2023examining} combine error messages from static analysis tools in the prompts\newtext{, NAVRepair~\cite{wang2024navrepair} combines AST node-type information and error types in prompts, Liu et al.~\cite{liu2024prompt} combines CWE-ID, vulnerability location, root cause in prompts to guide SPG}. Beyond software bugs, Ahmad~\cite{ahmad2023fixing} constructs prompts to fix hardware vulnerabilities.

\subsubject{\underline{\textit{B2. Few-shot prompting.}}} Few-shot prompting uses a few labeled examples. VQM~\cite{fu2024vision} provide example fixes of the CWE and data augumentation for repair. Though previously Fu et al.~\cite{fu2023chatgpt} tested \texttt{gpt-3.5-turbo} and \texttt{gpt-4} with three repair examples per prompt and found that it failed to generate correct patches for all vulnerable functions in their dataset.

\subsubject{\underline{\textit{B3. Chain-of-thought prompting.}}} Chain-of-thought (CoT) prompting improve LLMs' logic-based task performance by simulating human reasoning, which address the limitations of other prompting methods in improving reasoning capabilities.
\newtext{Nong et al.~\cite{nong2024chain}, Khan et al.~\cite{khan2024code}, VRpilot~\cite{kulsum2024case}, and ContractTinker~\cite{wang2024contracttinker} utilized CoT to break down AVR process into different reasoning steps (e.g., identify vulnerability types, vulnerability causes analysis), all finally integrate the results from previous reasoning results to guide SPG for these vulnerabilities. Also, VRpilot specify ``the fix should not break any functionality of the function'' in the prompt.}

\newtext{\subject{D. LLM Integration Approaches.} LLM integration approaches enable collaboration between LLMs and other tools beyond pure prompting to enhance AVR performance.} 

\newtext{\subsubject{\underline{\textit{D1. Multi-LLM Collaboration.} }} To overcome the incompleteness in single model approaches, VulMaster~\cite{zhou2024out} leverages two LLMs - CodeT5 as the fine-tunable backbone model and ChatGPT as the supplementary model for generating CWE examples and fixes, doubling the repair accuracy compared to approaches that rely solely on fine-tuned CodeT5.} 

\newtext{\subsubject{\underline{\textit{D2. LLM-External Tool Collaboration.} }}To enhance LLMs' ability to understand vulnerabilities (e.g., counterexamples), they are integrated with external tools. ESBMC-AI~\cite{tihanyi2023new} uses bounded model checker ESBMC~\cite{gadelha2018esbmc} to locate vulnerabilities and generate counterexamples (including stack traces, line numbers, and variable names), then feed the counterexamples and original code into LLMs for SPG, and iteratively verifies and refines the fixes through ESBMC until the code passes the verification. VRPilot~\cite{kulsum2024case} integrates external tools output (e.g., code sanitizers) into the prompt to guide SPG.}  

\takeaways{
\newtext{Learning-driven methods have transformed AVR by: (1) moving beyond traditional heuristic approaches and templates; (2) improving generalizability and offering more flexibility in inputs through fine-tuning, prompt engineering, and LLM integration approaches. Especially, prompts that provide contextual information (e.g., root causes) guide LLMs to generate more precise patches (e.g.,~\cite{kulsum2024case}); (3) LLM integration approaches enable complementary capabilities (e.g., different models and tools) to work together for precise analysis and iterative SPG, enhancing robustness.}
}

\openquestion{\begin{itemize}[leftmargin=*]
\item \textbf{\newtext{Component Dependency Analysis and Integration.}} 
Improving SPG requires \newtext{analyzing complex program dependencies and their security implications} ((e.g.,~\citep{pearce2023examining, wang2024navrepair, liu2024prompt, kulsum2024case})). \newtext{This includes: (1) understanding multi-component interactions like API calls and structure definitions, (2) tracking inter-procedural data flows, and (3) validating dependencies against security specifications. Thus, developing multimodal methods to analyze these may lead to more accurate, contextually informed repairs.}
    \item \textbf{Common Challenges in Learning-based Security Analysis.} Common problems like data sparsity, input length limitations, and training data availability in learning-based security analysis are even more challenging in AVR. For example, while fine-tuning has proven effective in AVR, the lack of high-quality \newtext{vulnerability repair} datasets(e.g.,~\cite{huang2022repairing})—comprising not only official patches but also semantically equivalent patches and clear evaluation metrics—remains a significant hurdle.
    \item \textbf{Generated Security Patches Selection.} 
    In non-deterministic models like GPT-4, \newtext{automatically validating and selecting the correct security patch from inconsistent outputs remains challenging. Unlike general bug fixes, security patches require complex analysis of exploitability and system-wide security implications, making automated validation methods still insufficient. This validation process currently relies heavily on human expertise (e.g.,~\cite{wu2023effective, khan2024code}).}
    \item \textbf{\newtext{Iterative AVR.}} \newtext{Tools feedback incorporation has proven AVR improvement (e.g.,~\cite{tihanyi2023new, kulsum2024case}), how can we leverage 
    % advanced 
    LLM integration usages (e.g., LLM multi-agents) with long memories to enhance it not only limited to statement/function?}
\end{itemize}}

\section{Benchmark Evaluation}
\label{sec:exp}

In Sections~\ref{sec:taxonomy_apr} and ~\ref{sec:patch_generation},we explore AVR advances and evaluate SPG methods. Here, our goal is to understand their practical strengths and weaknesses.

\subsection{Benchmark Evaluation Setup}

\subject{Dataset Selection.} We benchmark SPG methods by curating a dataset that spans various languages and weakness types. 
After reviewing vulnerabilities with fixes datasets (\cite{black2017sard, nikitopoulos2021crossvul, he2023large, bhandari2021:cvefixes,  ponta2019manually, fan2020ac,reis2021ground, shen2021localizing, gao2021beyond, bui2022vul4j,wu2023effective}),
we strategically select 4 datasets. 
% each chosen to fulfill a unique role in our research. 
Our selection criteria include: (1) prevalence in AVR research, (2) programming language diversity, (3) mix of synthetic and real-world vulnerabilities, and (4) real-world vulnerabilities test cases/exploits availability for verification. Finally, we selected: 
(1) synthetic dataset {\small$\mathcal{D}_{SARD}$}~\cite{black2018juliet} (used in learning-based AVR evaluation~\citep{pinconschi2022maestro, nong2024chain, islam2024code}). 
For our evaluation, we randomly sampled 1,000 samples from this dataset. (2) real-world Java vulnerabilities  {\small$\mathcal{D}_{Vul4J}$}~(\#79)~\cite{bui2022vul4j}, {\small $\mathcal{D}_{VJBench}$}~(\#42)~\cite{wu2023effective} and its transformations. 
(3) real-world C/C++ vulnerabilities {\small$\mathcal{D}_{ExtractFix}$}~(\#30)~\cite{gao2021beyond}, and {\small$\mathcal{D}_{Vulnloc}$}~(\#43)~\cite{shen2021localizing}, totaling 48 distinct vulnerabilities across {\small$\mathcal{D}_{ExtractFix}$} and {\small$\mathcal{D}_{Vulnloc}$}. 
See details in Appendix~\ref{app: data}.

\begin{table*}[htbp]
    % \vspace{-0.5 em}
    \caption{Evaluation results on C/C++ Vulnerability Benchmark. This table presents data in the format: Successful Repairs / Tests Passed / Successful Compilations / Total Tests Conducted (RSR). `NA' means the tool is not applicable to the benchmark.}
    \label{tab:cpp_benchmark_results}
    \centering
    {\small
    \resizebox*{0.7\textwidth}{!}{
        \begin{tabular}{|c|c|c|c|} 
            \hline
            Benchmark          & \multicolumn{1}{c|}{\#SARD}                 & \multicolumn{1}{c|}{\#ExtractFix} & \#VulnLoc  \\ 
            \hline\hline
            Senx~\cite{huang2019using}                & N/A                                          & 2/2/6/30(6.7\%)                              & 4/4/8/43(9.30\%)       \\
            VulnFix~\cite{zhang2022program}              & N/A                                          & 13/13/19/30(43.33\%)                             & \textbf{20/20/28/43(46.51\%)}      \\ 
            % \hline
            ExtractFix~\cite{gao2021beyond}         & N/A                                          & \textbf{21/21/25/30(70.00\%)}                             & 16/16/18/43(37.21\%)      \\ 
            \newtext{VRPilot~\cite{kulsum2024case}} & \newtext{N/A} & \newtext{17/17/22/30(56.67\%)}           &  \newtext{18/18/25/43(41.86\%)}           \\

            InCoder~\cite{fried2022incoder}            & 79/97/120/1000(7.90\%)                                     & 0/2/5/30(00\%)                               & 0/3/7/43(00\%)       \\ 
            % \hline
            Gemini-Pro~\cite{gemini-patching}            & {121/153}/971/1000(12.10\%)                                     & 1/1/25/30(3.33\%)                               & 0/0/32/43(00\%)       \\ 
            % \hline
            GPT-4-1106-preview~\cite{gpt_4} & \textbf{966/982/1000/1000(96.6\%)}                                    & 14/14/20/30(46.67\%)                             & 15/15/23/43(34.88\%)      \\ 
            \hline
            \end{tabular}}}
        % \vspace{-0. em}
\end{table*}

\begin{table*}[htbp]
    \caption{ Evaluation results on Java Vulnerability Benchmark. \textbf{VJBench-Trans(R)}:renaming transformation, \textbf{S}: structure change, \textbf{R+S}: changed by both.
     Data Format: Successful Repairs / Tests Passed / Successful Compilations / Total Tests Conducted (RSR). 
    }
    \label{tab:java_benchmark_results}
    \centering
    \small{
    \resizebox*{0.95\textwidth}{!}{
        \begin{tabular}{|c|c|c|c|c|c|} 
            \hline
            Benchmark          & \#VJBench          & \#VJBench-Trans(R) & \#VJBench-Trans(S) & \#VJBench-Trans(R+S) & \#VUL4J  \\%& \#Defects4J$^{NPE}$\\ 
            \hline\hline
            GPT-4-1106-preview~\cite{gpt_4} & 5/5/20/23(21.74\%) & 5/5/22/23(21.74\%) & 5/7/20/23(21.74\%) & 3/5/20/23(13.04\%)          & 26/30/56/70(37.14\%)  \\%& \textbf{12/12/17/18(66.67\%)}\\ 
            \newtext{VRPilot~\cite{kulsum2024case}} & \newtext{\textbf{6/9/20/23(26.08\%)}}         & \newtext{\textbf{6/9/22/23(26.08\%)}}         & \newtext{\textbf{6/8/20/23(26.08\%)}}          & \newtext{\textbf{5/6/20/23(21.74\%)}}           & \newtext{\textbf{29/31/53/70(41.43\%)}}      \\
            % \hline
            Gemini-Pro~\cite{gemini-patching}             & 2/2/14/23(8.70\%)              & 2/2/13/23(8.70\%)              & 2/2/14/23(8.70\%)              & 2/2/13/23(8.70\%)                & 7/10/38/70(10.00\%)            \\% & 5/5/11/18(27.78\%)   \\ 
            % \hline
            CodeT5~\cite{wang2021codet5}             & 0/0/0/23(00\%)           & 0/0/1/23(00\%)           & 0/0/0/23(00\%)           & 0/0/0/23(00\%)             & 2/2/6/70(2.86\%)         \\%&  1/1/8/18(5.56\%)  \\ 
            % \hline
            InCoder~\cite{fried2022incoder}            & 2/2/10/23(8.70\%)          & 1/1/6/23(4.35\%)           & 1/1/5/23(4.35\%)           & 1/1/6/23(4.35\%)             & 6/10/20/70(8.57\%)          \\%&  2/2/5/18(11.11\%)\\ 
            % \hline
            Fine-tuned-CodeT5~\cite{wu2023effective}  & 3/4/17/23(13.04\%)          & 3/3/17/23(13.04\%)          & 3/3/16/23(13.04\%)          & 2/2/18/23(8.70\%)            & 2/10/48/70(2.86\%)     \\%&   2/2/11/18(11.1\%)    \\ 
            % \hline
            Fine-tuned-InCoder~\cite{wu2023effective} & 3/4/15/23(13.04\%)          & 3/3/16/23(13.03\%)          & 4/4/16/23(17.39\%)          & 2/2/17/23(8.70\%)            & 6/12/51/70(8.57\%)       \\%&  2/2/7/18(11.11\%)   \\
            % \hline
            
            % VRPilot~\cite{kulsum2024case} & 5/6/20/23(21.74\%) & \textbf{5/5/22/23(21.74\%)} & \textbf{5/7/20/23(21.74\%)} & \textbf{5/7/20/23(21.74\%)} & \textbf{5/7/20/23(21.74\%)} & TBD \\
            % \hline
            % NPEFix~\cite{durieux2017dynamic} & NA & NA & NA & NA & NA & 1/7/11/18(5.56\%)\\
            \hline
            \end{tabular}}}
\end{table*}

\subject{Inclusion \& Exclusion Criteria.} 
 
We evaluate contemporary AVR approaches, including but not limited to work from premier conferences.
Tools were filtered based on artifact availability and must be: 
(1) be publicly available or accessible from authors, 
(2) executable, and 
(3) reproducible. Detailed artifacts status and selection criteria are discussed in Appendix~\ref{app:art}.

Ultimately, we evaluated \approachcnt tools, including property-guided approach Senx~\cite{huang2019using}, constraint approach ExtractFix~\cite{gao2021beyond}, combined approach (search + constraint) VulnFix~\cite{zhang2022program}, and learning-based methods, including code infilling models InCoder~\cite{fried2022incoder} and CodeT5~\cite{wang2021codet5}, their fine-tuned versions~\cite{jiang2023impact}, and generative models using prompt engineering, specifically \texttt{\small GPT4-1106-preview}~\cite{gpt_4} and \texttt{\small Gemeni-Pro}~\cite{gemini-pro}, \newtext{and VRPilot~\cite{kulsum2024case} which leverages CoT prompting with error messages, to ensure the fairness for evaluation, we also applied \texttt{\small GPT4-1106-preview} for VRPilot instead of \texttt{\small GPT-3.5-turbo} as claimed in their paper}. The experimental settings, tools applicability, and evaluation steps please refer to Appendix~\ref{sec:appendix-exp}.

\subject{Evaluation Metrics}. We evaluate the capability of SPG methods by measuring their repair success rate (RSR):

{
\scriptsize
\begin{equation}
RSR = \frac{\text{Number of Vulnerabilities Successfully Repaired}}{\text{Total Number of Vulnerabilities to be Repaired}}
\end{equation}
}

Note that for non-deterministic models, we generate 3 outputs for each input; as long as one of them passes the test, we include it in the calculation.

\subject{Evaluation Results.}
Tables~\ref{tab:cpp_benchmark_results} and \ref{tab:java_benchmark_results}  show the number of successful repairs, oracle (tests/exploits) passed, successful compilations, and total test counts. These metrics, along with RSR, highlight failure stages for each benchmark. We filtered Java datasets\footnote{If the same modification is applied across different files, we still count it as a single modification.} due to learning methods' limitations with multi-file modifications, retaining 23/42 vulnerabilities in { $\mathcal{D}_{VJBench}$} and 70/79 vulnerabilities in { $\mathcal{D}_{Vul4J}$}.
All real-world C/C++ benchmarks were kept since they could be tested on non-learning methods.

\subsection{Highlighted Findings}
% \todo{Change the subsection}
\label{subsec:conclusion}

As shown in Table~\ref{tab:cpp_benchmark_results} and Table~\ref{tab:java_benchmark_results}, AVR performance varies significantly across benchmarks.
\texttt{\small GPT-4-1106-preview} excelled on {\small $\mathcal{D}_{SARD}$} with an {\small RSR} of 96.6\%. However, its effectiveness dropped below 50\% in other benchmarks, highlighting its limitations with real-world vulnerabilities. Other learning-based models had {\small RSRs} under 20\%. In contrast, \texttt{VulnFix} and \texttt{ExtractFix} handled real-world scenarios better, though their RSRs were not exceed 70\%. \texttt{Senx}, limited to three vulnerability types, resulting in a low RSR on diverse benchmarks. \newtext{Additionally, while VRPilot demonstrates improved RSR when integrated with external tool outputs, its performance remains below that of non-learning-based methods.}

Learning-based methods show higher compilation rates than RSR but struggle with actual vulnerability fixes.
For instance, on {\small $\mathcal{D}_{Vul4J}$}, \texttt{\small GPT-4-1106-preview} compiled 56 patches with 30 passing test cases but only 26 correctly fixed vulnerabilities, highlighting insufficient test coverage. Despite improvements in compilation (0 to 17 compilable patches in {\small$\mathcal{D}_{VJBench}$}) with fine-tuned models (e.g., fine-tuned CodeT5), 
successful repairs remains low (0 to 3).

Due to insufficient repaired samples in {$\mathcal{D}_{VJBench}$} for meaningful comparisons,
we applied 3 mutation strategies: variable swapping, condition reconstruction, and loop transformation (Appendix~\ref{sec: appendix-mutate}) on {$\mathcal{D}_{SARD}$} with 200 random samples, still in high RSRs with \texttt{GPT-4-1106-preview} (97.5\%, 96.15\%, and 96.43\% respectively).

\finding{SPG methods show varying effectiveness across benchmarks with no consistently dominant approach.  The robustness of learning-driven method in our evaluation is excellent. \newtext{Iterative LLM-external tool integration enhances AVR.}}

\openrq{How can we improve learning-based methods to maintain high compilation rates while ensuring functional consistency and enhancing security? \newtext{Given commonly used RSR only measure the proportion of successful repairs, how can we develop comprehensive evaluation metrics that better reflect the real-world effectiveness (e.g., also consider the vulnerability severity)?}}

\begin{table}[h]
\centering
\caption{{Successful repairs by Scope of Change and Code Dependencies on $\mathcal{D}_{ExtractFix}$ and $\mathcal{D}_{Vulnloc}$}, \textbf{SH}: Single-hunk, \textbf{MH}: Multi-hunk, \textbf{Intra/Inter/Other}: corresponding dependencies}
\label{tab:complexity}
\resizebox*{0.48\textwidth}{!}{
\begin{tabular}{|>{\hspace{0pt}}m{0.32\linewidth}|>{\hspace{0pt}}m{0.135\linewidth}|>{\hspace{0pt}}m{0.135\linewidth}|>{\hspace{0pt}}m{0.106\linewidth}|>{\hspace{0pt}}m{0.106\linewidth}|>{\hspace{0pt}}m{0.106\linewidth}|} 
\hline
\multirow{2}{0.32\linewidth}{\diagbox[width=3.0cm]{Tool}{Factors}} & \multicolumn{2}{>{\hspace{0pt}}m{0.27\linewidth}|}{Scope of Change} & \multicolumn{3}{>{\hspace{0pt}}m{0.318\linewidth}|}{\small Code Dependencies}  \\ 
\cline{2-6}
                                                     & SH    & MH                                                          & Intra & Inter & Other                                                  \\ 
\hline \hline
\newtext{VRPilot}                                  & \newtext{13/25} & \newtext{9/23}                                                       & \newtext{8/15} & \newtext{7/23} & \newtext{7/10}                                                  \\ 
GPT-4-1106-prev                                   & 11/25 & 8/23                                                       & 8/15 & 5/23 & 6/10                                                   \\
% \hline
ExtractFix                                           & 15/25 & 6/23                                                       & 8/15 & 10/23 & 3/10                                                  \\ 
% \hline
VulnFix                                              & 12/25 & 8/23                                                       & 6/15 & 10/23 & 4/10                                                  \\
\hline
\end{tabular}}
% \vspace{-0.7 em}
\end{table}

% \subject{Factors.}
To understand which changes make vulnerabilities more likely to be repaired, we examine the following dimensions: \textit{\small Scope of Change}, and \textit{\small Code Dependencies}. Since benchmark $\mathcal{D}_{ExtractFix}$ and $\mathcal{D}_{Vulnloc}$ feature a broad application of various methods, our analysis primarily focuses on it. \textit{\small Scope of Change} refers to the number of modified code blocks, categorized into single-hunk (\newtext{modifications confined to one contiguous code block}) and multi-hunk changes (\newtext{changes across multiple separated blocks}). \textit{\small Code Dependencies} are analyzed at three levels: intra-procedural dependency (fixed entirely within a single function or method, without needing to involve other components or functions.), inter-procedural dependency (modifications within the vulnerable function involve components outside of this function), and others (e.g., built-in features).

Table~\ref{tab:complexity} indicates minimal impact of \textit{scope of changes} on repair success across methods. 
\newtext{VRPilot} shows worse performance (\newtext{RSR 30.43\%}) with inter-dependencies compared with ExtractFix and VulnFix.

\finding{“Scope of changes” is not a reliable metric to indicate the difficult level of AVR (although commonly used to assess the difficulty of repairing general bugs). In contrast, ``code dependencies'' have a more significant impact on RSR. }

\label{subject:complex}
\begin{lstlisting}[language=diff, style=diff, basicstyle=\ttfamily\scriptsize, caption={ Patch for \href{https://github.com/vadz/libtiff/commit/0a76a8c765c7b8327c59646284fa78c3c27e5490}{CVE-2017-7601}, which was not repaired by learning-based methods but by VulnFix},captionpos=b, label={lst:cve-2017-7601}]
@@ -1632,6 +1632,13 @@ JPEGSetupEncode(TIFF* tif)
        "Invalig horizontal/vertical sampling value");
        return (0);
    }
+   if( td->td_bitspersample > 16 )
+  {
+     TIFFErrorExt(tif->tif_clientdata, module,
+                   "BitsPerSample %d not allowed for JPEG",
+                   td->td_bitspersample);
+     return (0);
+  }
\end{lstlisting}
We further investigated vulnerabilities that could be repaired by non-learning-based but not by learning-based approaches. 
A closer look at the vulnerabilities that learning-based methods failed to repair reveals complex issues, including intricate control flow conditions, program dependencies, non-obtainable structure members, global variables, or specifications. 
\newtext{For example, in Listing~\ref{lst:cve-2017-7601}, the vulnerability is caused by shift operations with excessively large exponents, which is repaired by a single-hunk modification. This stems from a lack of input validation for \texttt{td->td\_bitspersample}, 
allowing values exceeding JPEG specification~\cite{itu_t81}'s 16-bit limit and causing integer overflow.
VulnFix and ExtractFix can apply the correct repair by computing correct constraints, e.g., VulnFix could generate the correct patch invariant at the crash point, which is further utilized for SPG, while learning-based methods miss details regarding the specification. However, if such specification is provided in prompts, \texttt{\small GPT-4-1106-preview} can correctly perform the repair.} We also analyzed the cases learning-based methods repaired but non-learning-based methods did not, and cases that both approaches repaired; see details in Appendix~\ref{app:case}.

\finding{Learning-based methods lack a comprehensive understanding of the entire program, struggling with vulnerabilities involving complex interrelations, atypical constructs, extensive program-wide constraints, or implicit constraints defined in specifications. In contrast, non-learning-based methods excel by leveraging the broader program context and computing critical constraints or invariants.}

\openrq{How can we systematically identify and extract vulnerability root causes to generate precise specifications that guide learning-based repairs? Additionally, how can we integrate contextual dependencies and derive constraints from textual sources (e.g., specification documents) to enhance AVR?}

 We analyzed the weakness different methods successfully fixed, focusing on performance across {\small CWE} categories.  For {\small $\mathcal{D}_{SARD}$}, \texttt{\small GPT-4-1106-preview} has achieved an {\small RSR} 100\% in 52/60 {\small CWEs}, likely due to the simple contextual information.

\begin{figure}[ht]
	\centering
 \subfigure[Repair on Real world C/C++ Benchmarks]{
        \includegraphics[width=0.95\linewidth]{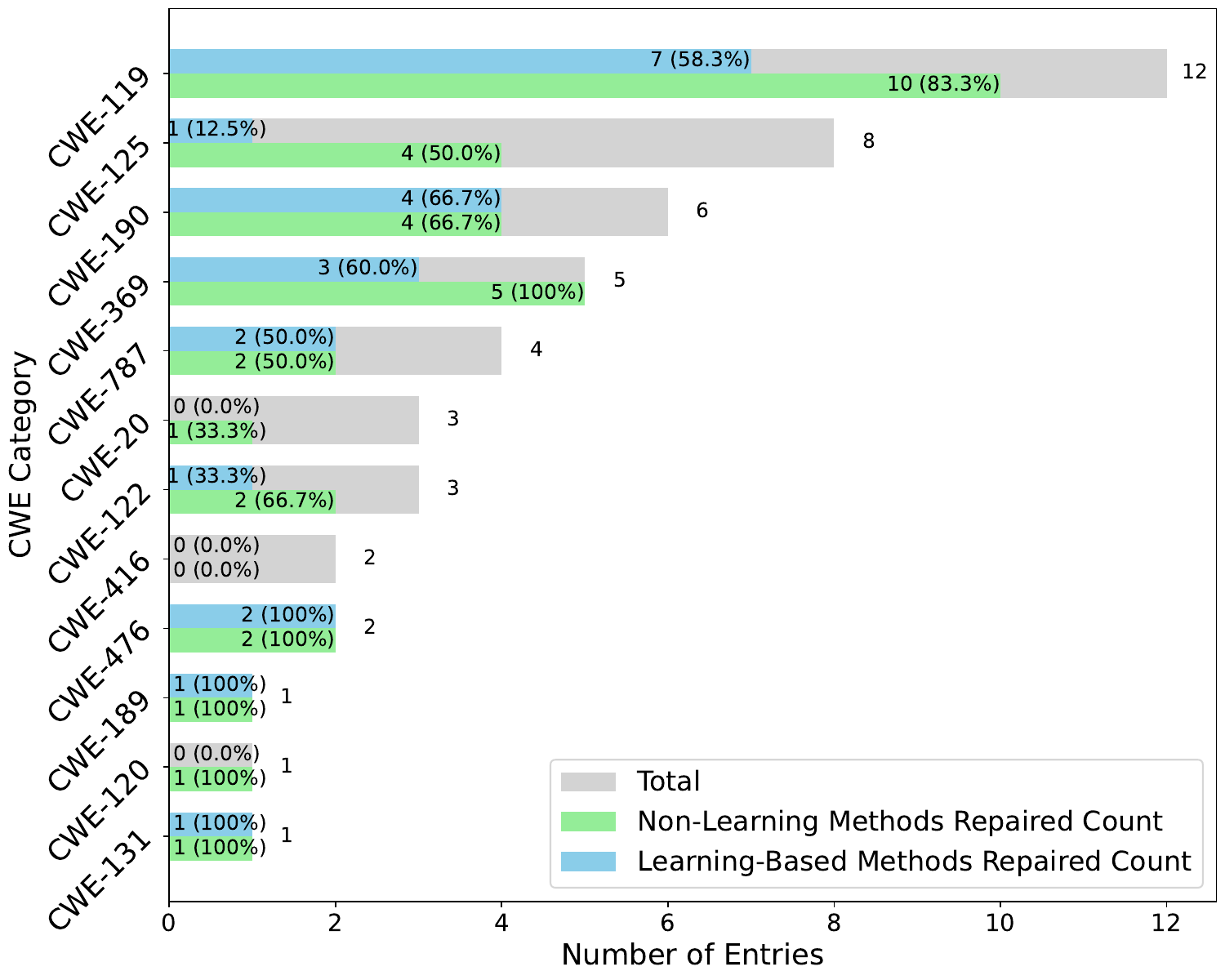}
        \label{fig:cbench}
        }
        
        \subfigure[Repair on Real world Java Benchmarks]{
	\includegraphics[width=0.95\linewidth]{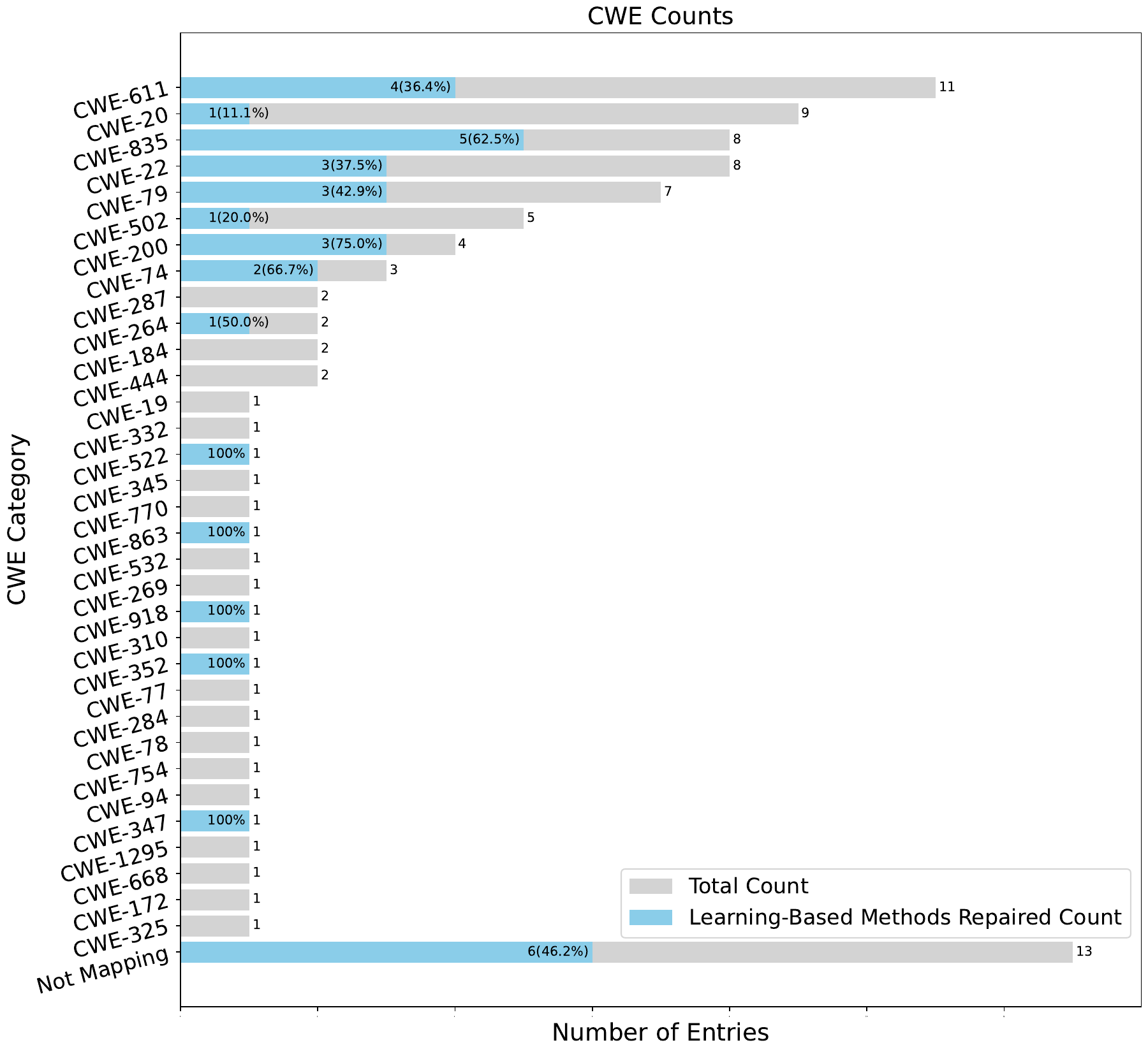}
        \label{fig:jbench}
 }
        \caption{Total vs. (Non-)Learning-Based Repair Counts Across CWE Categories in real-world benchmarks. }
        \label{fig:type}
        \vspace{-1em}
\end{figure}
 
 \label{subject:type}

Figure~\ref{fig:type} shows the repair distribution using learning and non-learning based approaches 
% } 
of CWE types across all real-world benchmarks. As shown in Figure~\ref{fig:cbench}, on each CWE type, no learning-based method performs better for any CWE. Specifically, both achieve 100\% RSR on CWE-476 ({NPDs}) and CWE-189 (Numeric Errors). For example, in the case of {CWE-189}, as seen in \texttt{\small CVE-2016-10094} (Listing ~\ref{lst:cve-2016-10094} in Appendix~\ref{app:case}), a numeric error was fixed by modifying the comparison from {\small \texttt{>=4}} to {\small \texttt{>4}} in the code, preventing heap-based overflow. For {CWE-476}, the main fix involved adding the necessary null check to prevent dereferencing null pointers. Figure~\ref{fig:jbench} shows that on each {CWE} with 2+ samples, the {RSR} is not greater than 75\%. Note that although some {CWEs} show high { RSR} , the limited number of data points makes it unreliable to draw any strong conclusions.

Moreover, benchmarks across different programming languages focus on distinct {CWE}  types, C/C++ benchmarks are primarily concerned with memory-related, like {CWE-119} ({Improper Restriction} of { Operations within Memory Buffer Bounds), while for Java, due to its garbage collections mechanism, tends to focus more on application-layer vulnerabilities (e.g., { CWE-611}, { Improper Restriction} of {XXE Reference}). The diverse application-layer vulnerabilities make it hard to design non-learning AVR tools for multiple CWEs in Java.

\finding{Learning-based AVR methods lag behind traditional approaches overall, though they excel at specific CWEs. The performance gap between real and synthetic datasets suggests synthetic data may inadequately evaluate AVR methods.}

\openrq{Can we establish a comprehensive benchmark for AVR? While some CWEs show high RSR, limited data makes strong conclusions unreliable.}

\section{Future Directions}
\vspace{-0.6em}

% In this section, we conclude by reflecting on the literature review in Section~\ref{sec:patch_generation} and the evaluation in Section~\ref{sec:exp}, to discuss possible directions for future research.
Based on our literature review (Section~\ref{sec:patch_generation}) and evaluation (Section~\ref{sec:exp}), we discuss future research directions here.

\subject{D1. Hybrid Approaches for SPG.} Based on Section~\ref{subsec:learning} and Section~\ref{subsec:conclusion}, 
%  Based on the challenges and open research questions illustrated in Section~\ref{subsec:learning} and Section~\ref{subsec:conclusion}, 
% while learning-based methods show promise in addressing vulnerabilities~\citep{pearce2023examining, wu2023effective}, a pressing challenge lies in effectively handling complex dependencies (Section~\ref{subsec:conclusion} Open RQ III). 
% Real-world vulnerabilities involve intricate dependencies across functions, variables, and files, which may overwhelm traditional program analysis due to issues like path explosion during symbolic execution. 
A hybrid approach that combining fuzzing, program analysis, etc. with LLMs for SPG may presents a promising direction. For example, while program analysis can identify critical paths and dependencies, it may struggle with scalability in large codebases. Here, LLMs can optimize the search space. This synergy between different techniques can be integrated into repair process. Also, fine-tuning LLMs with insights from program analysis may also helpful for context-aware and precise repair.  As discussed in Section~\ref{subject:complex}, artifacts like documentation and bug reports help with in-context learning. 
\newtext{Additionally, LLM could generate constraints or patch invariants based on the analysis of vulnerable and benign states. The effectiveness of LLMs for invariant generation has been demonstrated in recent works~\citep{wu2024llm, pirzada2024llm} but not tried in AVR yet.}
Compared to other software security tasks, hybrid approaches for AVR are more challenging as they must generate patches that account for dependencies, maintain semantic consistency, and avoid introducing new vulnerabilities.

\subject{D2. Domain-Specific AVR.} As discussed in Section~\ref{sec:taxonomy_apr}, while logic vulnerabilities receive less attention, they can be more severe, as privacy handling flaws may lead to data breaches and significant losses.  General-purpose AVR tools often struggle with logic vulnerabilities due to their intricate logic, which may involve subtle design flaws or complex program behavior that isn't as straightforward to detect or repair as memory-related issues.
 \newtext{Developing domain-specific methods for logic vulnerability repair, incorporating domain-specific languages, formal specification, and security properties and multi-modality information, is the foundation for reliable patches in domain-specific AVR (e.g., TAPFixer~\cite{yu2024tapfixer}). }

\subject{D3. AVR Benchmarks.}
\newtext{While numerous datasets on vulnerabilities and their fixes exist (Table~\ref{tab:data-details}), 
% most datasets primarily support ML-based vulnerability detection, which lacks 
most lack context and exploits. Current AVR datasets (Section~\ref{subject:type}) have limited CWE coverage. AVR needs comprehensive benchmarks with vulnerable code, patches, working exploits, and reproducible test environmentsm, challenging to create and maintain.%- a significant challenge to create and maintain.
}

\subject{D4. Interpretability of AVR.} As discussed in Section~\ref{subject:complex}, current AVR methods, especially learning-based ones, lack interpretability regarding which vulnerabilities are easier/harder to repair. Simple metrics like changed hunks or lines provide limited insights. More comprehensive metrics (control flow complexity, data flow changes, statement length) are necessary to improve interpretability of AVR performance.

\subject{D5. Automatic Generation of High-quality Specifications.} As discussed in Section~\ref{sec:patch_generation}, specifications are essential for extracting security properties and constraints. Section~\ref{subject:complex} further demonstrates that 
integrating more detailed information about vulnerable program points can enhance prompt engineering in learning-based methods, thereby increasing RSR. These insights highlight the need for better security-related specifications, as they define what constitutes a secure state and are crucial for enhancing the effectiveness of AVR. For example, in software documentation, security properties might be spread across different sections, using LLMs/NLP and automated reasoning techniques to extract and find potential violations may lead to security specific specifications.

\subject{D6. Verifier for Generated Security Patches.} 
As discussed in Section~\ref{sec:patch_generation}, 
existing approaches still need human intervention in SPV. When the search space is large (e.g., LLMs can generate multiple patches) potentially including a correct one. 
% Even for compilable patches, manual verification remains time-consuming.
% manual check will still be time-consuming, even though it is compilable. 
Also, incorrect patches can turn non-exploitable bugs into severe vulnerabilities\cite{wu2023mitigating} (e.g., the commit\cite{patch_dbfree} fixed a memory leak but introduced a new double-free vulnerability). To this end, we advocate for further developments in patch verification methods to improve AVR efficiency, which may related to formal verification techiniques.

\subject{\newtext{D7. LLM In AVR.}} \newtext{With the advancements in LLMs, LLMs might be used in any stage (VL, SPG, SPV) of AVR. As discussed in Section~\ref{sec: preliminaries} and Section~\ref{sec:patch_generation}, currently, the use of LLMs in these areas is mostly limited to analyzing the code itself or incorporating a small amount of vulnerability-specific information like vulnerability types. However, since LLMs are prone to hallucination which may lead to inaccurate outputs, LLMs for AVR cannot solely rely on the LLMs' capabilities themselves. It is crucial to integrate them with traditional complementary techniques. \textbf{For VL}, LLMs can help filter relevant statements from dependency graphs in slice-based VL (discussed in Section~\ref{sec: preliminaries}), reducing noise and improving the precision of the analysis. Moreover, LLMs can assist counterexample generation by generating inputs that expose differences in predicates thay may correlate with crashes. These counterexamples help rank predicates based on their likelihood of being root causes, improving the precision of localization. \textbf{For SPG}, as discussed in D1, LLM can be integrated with other techniques and further for patch synthesis in SPG. \textbf{For SPV}, though currently LLMs have been used for self-correction or self-checking~\cite{qiu2024correctbench}, these approaches alone cannot prove correctness. However, LLMs can serve multiple roles in enhancing verification approaches. First, LLMs can extract semantic properties and security invariants from patches, which can then be formally encoded as verification conditions for automated theorem provers or model checkers. Second, LLMs can assist in translating patches into intermediate verification languages or logical formulas that are more amenable to automated reasoning. Third, when verification fails, LLMs can help analyze counterexamples generated by theorem provers and suggest refinements to the verification conditions or patches.
This combination leverages LLMs' natural language understanding and code analysis capabilities while relying on the mathematical rigor of automated reasoning for formal verification, making SPV more robust. In whole AVR process, LLM agents serve to decompose complex repair tasks into actionable steps, interact with vulnerability analysis, and maintain critical context including code structure, repair history, etc. to facilitate effective AVR.}
Please refer to Appendix~\ref{app:future} for more future directions.

\section{Conclusion}
\label{sec:conclusion}
% This paper presents as systematization of knowledge about security patch generation methods in the process of vulnerability repair. After identifying and analyzing \papernumber representative papers from venues across security, software engineering, and programming language, we provide a detailed taxonomy, including characteristics, strengths of these approaches. We provide the open questions both theoretically and empirically, and reveals limitations. 

% This paper presents a systematization of knowledge concerning security patch generation methods in the process of vulnerability repair. After identifying and analyzing \papernumber representative papers from venues spanning security, software engineering, and programming languages, we offer a detailed taxonomy that includes the characteristics, strengths and weaknesses of these approaches. We also propose open research questions both theoretically and empirically.

We present an SoK concerning SPG in AVR, including a comprehensive taxonomy of their characteristics and trade-offs, with future research directions theoretically and empirically.
% . We offer a detailed taxonomy that includes the characteristics, strengths and weaknesses of these approaches. We also propose open research questions both theoretically and empirically.

%-------------------------------------------------------------------------------
% Reference
%-------------------------------------------------------------------------------

{
\small
{
\normalem
\bibliographystyle{unsrt}
\selectfont\bibliography{refs}}
}

\normalsize

%-------------------------------------------------------------------------------
% Appendix
%-------------------------------------------------------------------------------
% \appendix
% % \section*{Appendix}
% \section{Experimental Setting Details}

\appendix
\section*{Appendix}
\renewcommand{\thesubsection}{\Alph{subsection}}

The appendix illustrates more details on the experimental setting, dataset selection, mutation strategies, security patch generation artifacts status, and case studies.
% \faysal{Ensure that we have well document ted Appendix.}

%-------------------------------------------------------------------------------
% \subsection{Figures}
% \label{app:fig}
% \ying{This figure needs discuss} \ying{Could be deleted}
% \begin{figure}[H]
%     \centering
%     \subfigure[The original vulnerability of VJBench-Halo-1(Directory Traversal)]{\label{fig:a}\includegraphics[width=0.45\textwidth]{Figs/cases/original_halo1.pdf}}

%     \subfigure[VJBench-Halo-1 transformed by renaming]{\label{fig:a}\includegraphics[width=0.45\textwidth]{Figs/cases/rename_halo1.pdf}}

%     \caption{Halo-1 before and after transformation by renaming(i.e., replace variables, function names, etc. with synonyms)}
%     \label{fig:rename_case}
% \end{figure}
% \begin{appendices}
\subsection{Dataset Selection Details}
\label{app: data}
\vspace{-0.1cm}

We selected the vulnerability and fixes dataset from the known datasets shown in Table~\ref{tab:data-details}. In these datasets, only SARD is synthetic data, only VJBench\&VJbench-Trans, Vul4J, VulnLoc, ExtractFix consist test suites/exploits, so we select these datasets as our benchmarks.

\begin{table}[H]
\caption{Vulnerability \& Fixes Dataset Details}
\label{tab:data-details}
\centering
\begin{threeparttable}
\resizebox*{0.48\textwidth}{!}{
\begin{tblr}{|c|c|c|c|c|c|c|}%{
%   % colspec = {Q[260]Q[171]Q[73]Q[85]Q[75]Q[98]Q[162]},
%   hlines,
%   cells = {c}
% }
\hline
\textbf{Name}                                        & \textbf{Programming Language}~\tnote{1}      & \textbf{Data Type}~\tnote{2} & \textbf{CWE}~\tnote{3} & \textbf{Total}~\tnote{4} & \textbf{Patch}~\tnote{5} & \textbf{Test Suites/Exploit}~\tnote{6} \\
\hline\hline
SVEN\cite{he2023large}                      & Python\&C                 & R         & 9         & 1,606      & \CheckmarkBold              & \XSolidBrush                            \\
\hline
CVEfixes\cite{bhandari2021:cvefixes}        & \(>\)30                        & R         & 209       & 5,365      & \CheckmarkBold              & \XSolidBrush                            \\
\hline
SARD\cite{black2017sard}                    & C/C++                     & S         & 60        & 25,297     & \CheckmarkBold              & \XSolidBrush                           \\
\hline
Big-Vul\cite{fan2020ac}                     & C/C++                     & R         & 91        & 3,754      & \CheckmarkBold              & \XSolidBrush                            \\
\hline
CrossVul\cite{nikitopoulos2021crossvul}     &  \(>\)40 & R         & 161       & 5,131      & \CheckmarkBold              & \XSolidBrush                            \\
\hline
% Ground-truth\cite{reis2021ground}           & \(>\)20              & R         &    \todo{XX}       &   \todo{XX}         & \CheckmarkBold              & \XSolidBrush                            \\
% \hline
Vulas\cite{ponta2019manually}               & Java                      & R         & NA        & 624        & \XSolidBrush              & \XSolidBrush                            \\
\hline
VJBench\&VJBenchTrans\cite{wu2023effective} & Java                      & R         & 23        & 42         & \CheckmarkBold              & \CheckmarkBold                            \\
\hline
Vul4J\cite{bui2022vul4j}                    & Java                      & R         & 23        & 79         & \CheckmarkBold              & \CheckmarkBold                            \\
\hline
VulnLoc\cite{shen2021localizing}            & C/C++                     & R         &    12       & 43         & \CheckmarkBold              & \CheckmarkBold                            \\
\hline
ExtractFix\cite{gao2021beyond}              & C/C++                     & R         &     11      & 30         & \CheckmarkBold              & \CheckmarkBold                            \\
\hline
SecBench\cite{reis2017secbench}             & \(>\)20                        & R         &  51        & 676        & \XSolidBrush              & \XSolidBrush    \\
\hline
\end{tblr}}
\begin{tablenotes}
\scriptsize \item [1] { Programming Language: Indicates the types of programming languages covered by each dataset. }
 \item [2]  Data Type: `R' stands for real-world vulnerability and `S' signifies synthetic data. 
\item [3] CWE: The number of CWE covered by the dataset. `NA' signifies data not provided. 
\item [4] Total: The total number of entries or cases in the dataset. 
\item [5] Patch: Indicates whether patches are included in the dataset (\CheckmarkBold / \XSolidBrush). 
\item [6] Test Suites/Exploit: Whether test suites or exploits are available (\CheckmarkBold / \XSolidBrush).
% \footnotesize
% \item [1] Programming Language means the dataset contains what kinds of 
% \item [2] Note 2
\end{tablenotes}
\end{threeparttable}
\end{table}

\begin{table*}[!ht]
\centering
\caption{Artifacts Status}
\begin{threeparttable}
\resizebox*{\textwidth}{!}{
\SetTblrInner{rowsep=0pt}
\begin{tabular}{|c|c|c|c|c|c|c|c|} 
\hline
\textbf{Tool}         & \textbf{Link}                                                             & \textbf{Category}         & \textbf{Language}~\tnote{1} & \textbf{Accessible}~\tnote{2} & \textbf{Instruction}~\tnote{3} & \textbf{Executable}~\tnote{4} & \textbf{Reproducible}~\tnote{5}  \\ 
\hline\hline
ErrDoc~\cite{tian2017automatically}       & \url{https://github.com/yuchi1989/ErrDoc}                        & Template-Guided   & C/C++    & \CheckmarkBold          & \halfcirc  &  /          & /             \\ 
\hline
NPEFix~\cite{durieux2017dynamic}       & \url{https://github.com/SpoonLabs/npefix}                        & Template-Guided   & Java     & \CheckmarkBold          & \fullcirc           & \CheckmarkBold          & \CheckmarkBold             \\ 
\hline
BovInspector~\cite{gao2016bovinspector} & \url{https://github.com/BovInspector/project}                    & Template-Guided   & C/C++    & \CheckmarkBold          & \halfcirc  & /          & /             \\ 
\hline
SAVER~\cite{hong2020saver}        & \url{https://github.com/kupl/SAVER_public/}                      & Template-Guided   & C/C++    & \XSolidBrush          & \fullcirc           & /          & /             \\ 
\hline
IntPTI~\cite{cheng2019automatic}       & \url{https://github.com/45258E9F/IntPTI}                         & Template-Guided   & C/C++    & \CheckmarkBold          & \fullcirc           & \XSolidBrush          & /             \\ 
\hline
SGuard~\cite{nguyen2021sguard}       & \url{https://github.com/duytai/sGuard}                           & Template-Guided   & Solidity & \CheckmarkBold          & \halfcirc & /          & /             \\ 
\hline
Elysium~\cite{ferreira2022elysium}      & \url{https://github.com/christoftorres/Elysium}                  & Template-Guided   & Solidity & \CheckmarkBold          & \fullcirc           & \CheckmarkBold          & \CheckmarkBold             \\ 
\hline
EVMPatch~\cite{rodler2021evmpatch}     & \url{https://github.com/uni-due-syssec/evmpatch-developer-study} & Template-Guided   & Solidity & \CheckmarkBold          & \fullcirc           & \CheckmarkBold          & \CheckmarkBold             \\ 
\hline
Senx~\cite{huang2019using}       & Ask from authors                                                 & Template-Guided   & C/C++    & \CheckmarkBold          & \fullcirc           & \CheckmarkBold          & \CheckmarkBold             \\ 
\hline
IntRepair~\cite{muntean2019intrepair}    & \url{https://github.com/TeamVault/IntRepair}                     & Template-Guided   & C/C++    & \CheckmarkBold          & \halfcirc  & /          & /             \\ 
\hline
LeakFix~\cite{gao2015safe}      & \url{https://sei.pku.edu.cn/gaoqing11/leakfix}                   & Template-Guided   & C/C++    & \XSolidBrush          & \emptycirc & /          & /             \\ 
\hline
Genprog~\cite{le2011genprog}      & \url{https://github.com/squaresLab/genprog-code}                 & Search-Based     & C/C++    & \CheckmarkBold          & \fullcirc           & \CheckmarkBold          & \CheckmarkBold             \\ 
\hline
Tan et al.~\cite{tan2016anti} & \url{https://anti-patterns.github.io/search-based-repair/}       & Search-Based     & C/C++    & \CheckmarkBold          &  \emptycirc          & /          & /             \\ 
\hline
DeFinery~\cite{tolmach2022property}     & \url{https://github.com/palinatolmach/DeFinery}                  & Search-Based     & Solidity & \CheckmarkBold          & \fullcirc           & \CheckmarkBold          & \CheckmarkBold             \\ 
\hline
FixMorph~\cite{shariffdeen2021automated}     & \url{https://fixmorph.github.io/}                                & Search-Based     & C/C++    & \CheckmarkBold          & \fullcirc           & \CheckmarkBold          & \CheckmarkBold             \\ 
\hline
PeachWeave~\cite{shariffdeen2020automated}   & \url{https://github.com/rshariffdeen/PatchWeave}                 & Search-Based     & C/C++    & \CheckmarkBold          &  \fullcirc          & \CheckmarkBold          & \CheckmarkBold            \\ 
\hline
FootPatch~\cite{van2018static}    & \url{https://github.com/squaresLab/footpatch}                    & Search-Based     & C/C++    & \CheckmarkBold          & \fullcirc            & \CheckmarkBold          & \CheckmarkBold             \\ 
\hline
VulnFix~\cite{zhang2022program}      & \url{https://github.com/yuntongzhang/vulnfix}                    & Constraint-Based & C/C++    & \CheckmarkBold          & \fullcirc            & \CheckmarkBold          & \CheckmarkBold             \\ 
\hline
SemFix~\cite{nguyen2013semfix}       & \url{https://github.com/QIANZECHANG/SemFix}                      & Constraint-Based & C/C++    & \CheckmarkBold          & \emptycirc           & /          & /             \\ 
\hline
Memfix~\cite{lee2018memfix}       & \url{http://prl.korea.ac.kr/MemFix}                              & Constraint-Based & C/C++    & \XSolidBrush          & \emptycirc & /          & /             \\ 
\hline
CPR~\cite{shariffdeen2021concolic}          & \url{https://cpr-tool.github.io/}                                & Constraint-Based & C/C++    & \CheckmarkBold          & \fullcirc            & \CheckmarkBold          & \CheckmarkBold             \\ 
\hline
ExtractFix~\cite{gao2021beyond}   & \url{https://extractfix.github.io/}                              & Constraint-Based & C/C++    & \CheckmarkBold          & \fullcirc            & \CheckmarkBold          & \CheckmarkBold             \\ 
\hline
Symlog~\cite{liu2023program}       & \url{https://github.com/symlog/symlog}                           & Constraint-Based & /        & \CheckmarkBold          & \halfcirc  & /          & /   \\ 
\hline
Nopol~\cite{xuan2016nopol}        & \url{https://github.com/SpoonLabs/nopol/}                        & Constraint-Based & Java     & \CheckmarkBold          & \fullcirc            & \CheckmarkBold          & \CheckmarkBold             \\ 
\hline
Chida et al.~\cite{chida2022repairing} & \url{https://github.com/NariyoshiChida/SP2022?}                  & Constraint-Based & /        & \CheckmarkBold          & \emptycirc & /          & /             \\ 
\hline
LLM-Vul~\cite{wu2023effective}    & \url{https://github.com/lin-tan/llm-vul}                         & Learning-Based   & Java     & \CheckmarkBold          & \fullcirc            & \CheckmarkBold          & \CheckmarkBold            \\ 
\hline
VulRepair~\cite{fu2022vulrepair}    & \url{https://github.com/awsm-research/VulRepair}                 & Learning-Based   & C/C++    & \CheckmarkBold          & \halfcirc            & /          & /             \\ 
\hline
VRepair~\cite{chen2022neural}      & \url{https://github.com/ASSERT-KTH/VRepair}                      & Learning-Based   & C/C++    & \CheckmarkBold          & \halfcirc           & /          & /             \\ 
\hline
SeqTrans~\cite{chi2022seqtrans}     & \url{https://github.com/chijianlei/SeqTrans}                     & Learning-Based   & C/C++    & \CheckmarkBold          & \halfcirc  & /          & /             \\
\hline
VRPilot~\cite{kulsum2024case}     & \url{http://tinyurl.com/vrpilot-artifacts}                     & Learning-Based   & C/C++/Java    & \CheckmarkBold          & \fullcirc  & \CheckmarkBold          & \CheckmarkBold              \\
\hline
VulMaster~\cite{zhou2024out}     & \url{https://github.com/soarsmu/VulMaster_}                     & Learning-Based   & C/C++/Java    & \CheckmarkBold          & \halfcirc  & /          & /              \\
\hline
ContractTinker~\cite{wang2024contracttinker}     & \url{https://github.com/CheWang09/LLM4SMAPR}                     & Learning-Based   & Solidity    & \CheckmarkBold          & \fullcirc  & \CheckmarkBold          & \CheckmarkBold              \\
\hline
\end{tabular}}
\begin{tablenotes}
    \scriptsize{
    \item [1] The programming language tested in the paper; "/" indicates the paper illustrates its applicability to multiple programming languages.
    \item [2] Accessible: The artifact is available for use.
    \item [3] Instruction: The artifact includes clear instructions (covering environment, dependencies, versions, etc.) for all modules. \\ \fullcirc indicates the artifact has very clear instructions. \\ \halfcirc indicates the artifact has instructions but not for all modules or does not have the code for all modules(e.g., lack the module for data preprocessing). \\ \emptycirc indicates the artifact lacks instructions.
    \item [4] Executable: The tool can be successfully executed by following the provided instructions.
    \item [5] Reproducible: The results can be reproduced using the tool.}
\end{tablenotes}
\end{threeparttable}
\label{tab:artifact}
\vspace{-0.2cm}
\end{table*}

The datasets we selected are:
\vspace{-4pt}

\begin{itemize}[itemsep=-0.5pt]
    \item[(i)] \textbf{Synthetic Dataset}, $\mathcal{D}_{SARD}$~\cite{black2018juliet}: The SARD Dataset's Juliet C/C++ includes a total of 25,297 data points. We randomly sampled 1,000 instances(across 60 CWEs) from this comprehensive dataset. $\mathcal{D}_{SARD}$ is widely used in learning-based AVR works(e.g., ~\citep{pinconschi2022maestro, nong2024chain, islam2024code}).
    \item[(ii)] \textbf{Real-World Vulnerabilities in Java}, $\mathcal{D}_{Vul4J}$~\cite{bui2022vul4j},  $\mathcal{D}_{VJBench}$ and its transformations~\cite{wu2023effective}: The dataset includes 79 reproducible Java vulnerabilities across 25 CWEs in Vul4J~\cite{bui2022vul4j}, and 42 reproducible vulnerabilities in VJBench and VJBench-trans, which apply transformations like identifier renaming and code structure changes. 
    % To address the lack of NPD bugs, which are a leading security issue in production environments, we have also included 18 NPDs from Defects4J~\cite{just2014defects4j}.
    All these datasets include test cases, providing deep insights into Java vulnerabilities.
    \item[(iii)] \textbf{Real-World Vulnerabilities in C/C++}, $\mathcal{D}_{ExtractFix}$ and $\mathcal{D}_{Vulnloc}$~\cite{gao2021beyond, shen2021localizing}: This dataset contains vulnerabilities from ExtractFix(30)~\cite{gao2021beyond}, Vulnloc(43)~\cite{shen2021localizing} benchmark, as well as corresponding exploits. Specifically, it encompasses 48 real-world unique vulnerabilities from various projects, such as Binutils\cite{binutils} and Libtiff~\cite{libtiff}.  
\end{itemize}

Also, for the synthetic data $\mathcal{D}_{SARD}$, where defects and fixes are often associated with \texttt{callee} functions, caller functions' naming conventions, such as \texttt{Goodxxx} or \texttt{Badxxx}, may lead LLMs to erroneously transform ``good" to  ``bad". To address this issue, we implemented an inter-procedural analysis with a depth of one, embedding \texttt{callee} functions' bodies and parameters within their \texttt{caller} functions(As shown in Figure~\ref{fig:preprocess}). For other benchmarks, the original input structure is preserved because the function does not exist such naming issues in real-world vulnerabilities; also, such inter-procedural processing on real-world vulnerabilities could cause the function to too long, which exceeds the limited token of current learning-based methods.

\begin{figure}[H]
\vspace{-0.3cm}
    \centering
\includegraphics[scale=0.35]{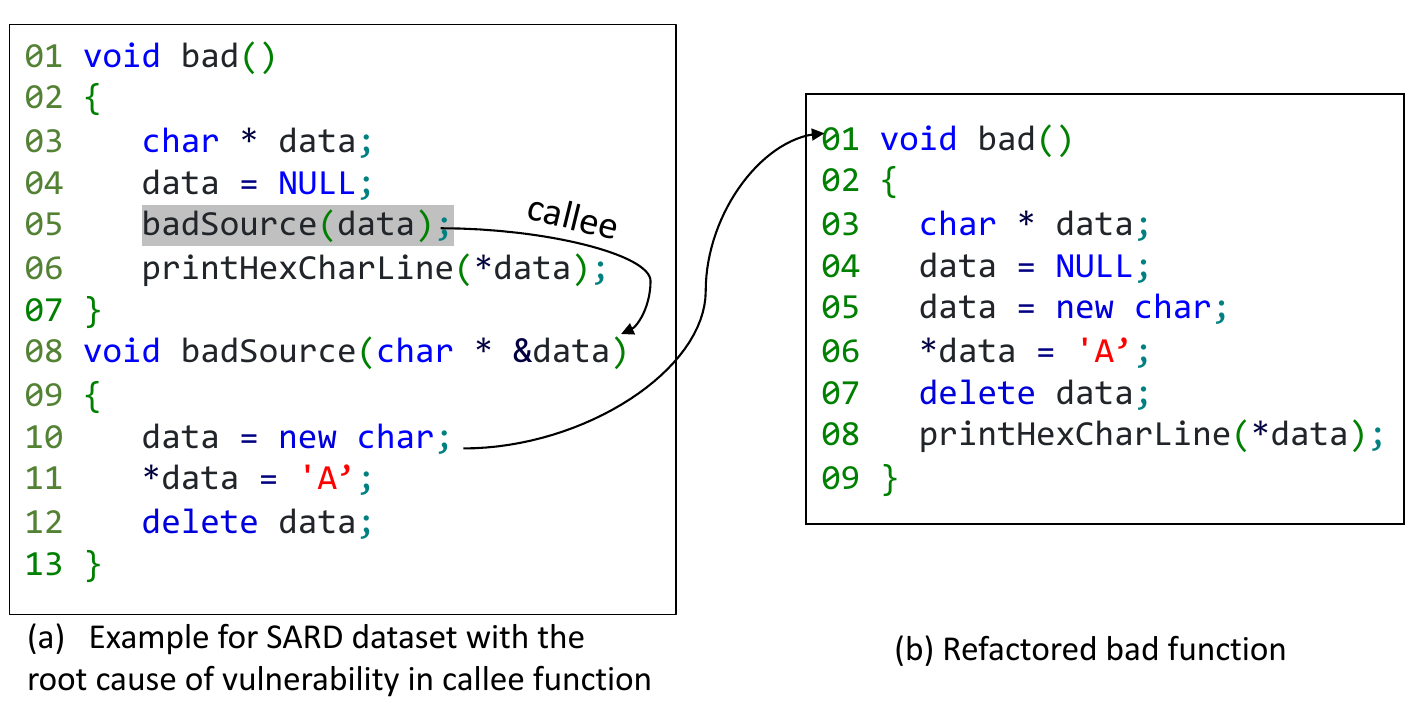}
    \caption{Inter-procedural processing for $\mathcal{D}_{SARD}$}
    \label{fig:preprocess}
    \vspace{-0.15cm}
\end{figure}
\subsection{Security Patch Generation Artifacts.}
\vspace{-0.1cm}
\label{app:art}

As shown in Table~\ref{tab:artifact}, we have summarized the artifact information that the authors mentioned as open source in their papers(Although some tools are not specifically designed for security vulnerabilities, they have mentioned or provided cases addressing security vulnerabilities in their experiments, e.g., GenProg~\cite{le2011genprog}). To select the tools, the basic requirements are: 1) publicly available or obtainable from the authors, 2) clear instructions for running the tool, 3) executable, and 4) reproducible results. For obtainable from authors, we only request tools from authors who have been compared in other papers and are stated as open-source in their papers. Among \newtext{\artifactcnt} tools, \newtext{16 (50.00\%)} satisfy all these criteria (Note that due to randomness of LLM, we only require comparable results instead of totally same results). However, we exclude the following tools from our experiment:
\begin{itemize}
    \item FixMorph, PeachWeave, and FootPatch: These 3 tools target the task of patch backporting, while this paper target general SPG instead of backporting.
    \item Elysium, Definery and ContractTinker: These three tools target the programming language Solidity, which is not included in the dataset.
    \item Nopol and NPEFix: These two tools focus on Null Pointer Dereferences(NPDs) in Java, whereas the Java benchmark did not consist NPDs.
    \item GenProg: This tool requires positive and negative test cases in C/C++, while the selected two real-world C/C++ datasets only provide one exploit without additional positive test cases.
\end{itemize}

Finally, in our test, we used the following tools: Senx~\cite{huang2019using}, VulnFix~\cite{zhang2022program}, ExtractFix~\cite{gao2021beyond}, LLM-Vul~\cite{wu2023effective}, \newtext{VRPilot~\cite{kulsum2024case}}. Note that in LLM-Vul, the authors compare different large models of vulnerability repair. We included the selection of CodeT5, InCoder, and their fine-tuned models to better compare performance before and after fine-tuning. For Codex, OpenAI recommends using GPT-3.5+ to replace Codex, so in our experiment, we used \texttt{GPT-4-1106-preview}. Additionally, we also tested Google's Gemini-Pro~\cite{gemini-patching}.

% \recprac{Following our evaluation and tries on AVR artifacts, we propose several practice recommendations:
% \begin{itemize}
%     \item Ensure that the open-source tool is accessible as described in the paper.
%     \item Provide detailed instructions for each step to ensure that reproducible experiments conducted by others are fair and consistent.
%     \item Include all necessary modules in the artifacts, as variations in any module (e.g., preprocessing) can significantly impact results and potentially lead to unfair comparisons. Building a Docker image for the tool or integrated tools is recommended as a good practice (e.g., Cerberus~\cite{shariffdeen2023cerberus}).
% \end{itemize}}

% \end{appendices}

\subsection{Experimental Setting Details}
% \vspace{-0.1cm}
\label{sec:appendix-exp}
\subject{Experimental Environment Settings.} Our experiments were conducted on Ubuntu 22.04.1 X86\_64. To handle the varied dependencies needed for reproducing vulnerabilities, we employed Docker for consistent environments. We accessed \texttt{GPT-4-1106-preview} and \texttt{Gemini-Pro} via APIs with prompt engineering, and used the fine-tuned CodeT5 and InCoder models from Huggingface~\cite{huggingface}, as used by Wu et al.~\cite{wu2023effective}. 
We used Joern (v2.0.110)~\cite{joern} for $\mathcal{D}_{SARD}$ preprocessing, CodeQL (v2.15.1) to detect remaining weakness. 
% For details on experimental parameters, prompts, and model settings, see Appendix~\ref{sec:appendix-exp}.

\subject{Tools Applicability.} Note that not all methods are applicable across benchmarks due to varying requirements (e.g., programming language) and target datasets. For instance, Vulnfix and ExtractFix require an exploit to trigger the vulnerability, \newtext{VRPilot needs exploit/test case feedback,} so we applied these methods only on $\mathcal{D}_{ExtractFix\&Vulnloc}$. CodeT5 and its fine-tuned model were applied exclusively to Java benchmarks, as their training data lacked C/C++ datasets\cite{jiang2023impact}.

\subject{Evaluation Steps.} We assessed the correctness of generated security patches through a three-step process.
First, we ensured the patched code compiled successfully; failure indicated an incorrect patch.  
Second, we used an oracle mechanism: patches for vulnerabilities with existing tests (all real-world vulnerabilities) were verified against those tests, while synthetic data ($\mathcal{D}_{SARD}$) were checked using CodeQL~\cite{codeql}. Finally, all generated patches were manually reviewed for semantic equivalence, functionality integrity, and absence of new vulnerabilities. To reduce subjectivity, two authors independently reviewed 200 randomly selected patches, achieving an agreement rate of \agreerate and a Cohen's kappa of \chkappa. 

\textbf{Input Configuration.} For learning-based methods, we summarized the input configuration from related literature~\cite{wu2023effective, pearce2023examining, nong2024chain, wu2023exploring} and the models' usages. 
Among these models, for deterministic models(i.e., the models that produce the same output for the same input, including CodeT5, InCoder, and their fine-tuned models), we only run them once; for other non-deterministic models(i.e., the models that may produce different outputs for the same input, including GPT4 and Gemini here), we run them three times, as long as one of the generated patches is correct, we regard them as correct. 
Especially, the practice prompts are derived from the above studies, though Nong et al.~\cite{nong2024chain} highlighted the efficacy of CoT prompting. However, in the step of SPG, its core still lies in providing LLMs with the weakness information and the vulnerability location. So we keep providing CWE ID, CWE Name, and the buggy location for such generative models. For non-learning-based methods, we deploy the docker or virtual machine that the authors provided. The detailed input formats of learning-based methods are shown in Table ~\ref{tab:learning-config}.
\begin{table}[!ht]
\centering
\caption{Input configuration of learning-based methods}
\resizebox*{0.47\textwidth}{!}{
\begin{tabular}{p{0.8cm}|p{1.6cm}|p{5.0cm}} 
\hline
\textbf{Model}              & \textbf{Version}           & \textbf{Input}                                  \\ 
\hline
{\small GPT4   }            & {\small \texttt{gpt4-1106-} \texttt{preview}} & {\scriptsize \{``system": ``You are a security patch generator. You will be given a vulnerable code, the CWE ID of it is \{CWE\_ID\}, i.e.,\{CWE\_NAME\}, the buggy line has been commented with "/*BUG*/"please directly provide the fixed code without any explanation.", ``user": vulnerable function\} }              \\ 
\hline
{\small Gemini  }           & {\small \texttt{Gemini-pro} }       & {\scriptsize \{"system\_instruction": "You are a security patch generator. You will be given a vulnerable code, the CWE ID of it is \{CWE\_ID\}, i.e.,\{CWE\_NAME\}, the buggy line has been commented with "/*BUG*/"please directly provide the fixed code without any explanation.", "user": vulnerable function\}}  \\ 
\hline
{\small CodeT5}             &{\footnotesize \texttt{codet5-large} }     & {\scriptsize Mask buggy lines with $\langle extra\_id\_i \rangle$and input the vulnerable function(i depends on the hunks(n) of modified code in the official patch, from 0 to n ). In the input vulnerable function, also comment on the buggy lines.}                                                                                        \\ 
\hline
{\small InCoder}            & {\small \texttt{incoder-6B}  }      & {\scriptsize Mask buggy lines with and input the vulnerable function, (i depends on the hunks(n) of modified code in the official patch, from 0 to n ). In the input vulnerable function, also comment on the buggy lines.}                                                                                            \\ 
\hline
{\small Fine-tuned CodeT5}  &        -           & {\scriptsize The same with CodeT5.}                                                                                                                                                                                                                                                                                    \\ 
\hline
{\small Fine-tuned InCoder} &          -         & {\scriptsize The same with InCoder.}                                                                                                                                                                                                                                                                                    \\
\hline
\end{tabular}}
\label{tab:learning-config}
\end{table}

\subsection{Mutation Strategies}
\vspace{-0.15cm}

\label{sec: appendix-mutate}
To measure robustness, we test three mutation strategies:
\begin{itemize}
    \item \textbf{Variable swapping}: This mutation strategy involves identifying variables within a function and shuffling their names. If only one variable exists, it is replaced with a randomly generated string of the same length. 
    \item \textbf{Condition structure reconstruction}: This mutation strategy changes the \texttt{if-else} branch, for example, change   {\tt \small if(a) \{BlockA\} else \{BlockB\}} to {\tt \small if(!a) \{BlockB\} else \{BlockA\}}.
    \item \textbf{Loop transformation}: Transform the loop constructs(e.g., {\tt \small for,\ while,\ do-while}) into a different type of loop structure.
\end{itemize}

\vspace{-0.15cm}

\subsection{Additional Case Studies}
\label{app:case}
\begin{lstlisting}[language=diff, style=diff, basicstyle=\ttfamily\scriptsize, caption={ Patch for Gnubug-26545, which was repaired by learning-based methods but not by non-learning based methods}, captionpos=b, label={lst:gnubug-26545}]
@@ -287,7 +287,7 @@ fillpattern (int type, unsigned char *r, size_t size)
    r[0] = (bits >> 4) & 255;
    r[1] = (bits >> 8) & 255;
    r[2] = bits & 255;
-   for (i = 3; i < size / 2; i *= 2)
+   for (i = 3; i <= size / 2; i *= 2)
      memcpy (r + i, r, i);
    if (i < size)
      memcpy (r + i, r, size - i);
\end{lstlisting}

\begin{lstlisting}[language=diff, style=diff, basicstyle=\ttfamily\scriptsize, caption={ Patch for CVE-2016-10094, which was repaired by both learning-based methods and non-learning based methods}, captionpos=b, label={lst:cve-2016-10094}]
@@ -2895,7 +2895,7 @@ tsize_t t2p_readwrite_pdf_image_tile(T2P* t2p, TIFF* input, TIFF* output, ttile_
			return(0);
		}
		if(TIFFGetField(input, TIFFTAG_JPEGTABLES, &count, &jpt) != 0) {
-			if (count >= 4) {
+			if (count > 4) {
                int retTIFFReadRawTile;
        /* Ignore EOI marker of JpegTables */
				_TIFFmemcpy(buffer, jpt, count - 2);
\end{lstlisting}

\begin{lstlisting}[language=diff, style=diff, basicstyle=\ttfamily\scriptsize, caption={ Patch for CVE-2016-9273, which was not repaired by both learning-based methods and non-learning based methods}, captionpos=b, label={lst:cve-2016-9273}]
@@ -5,6 +7,15 @@ TIFFNumberOfStrips(TIFF* tif)
	TIFFDirectory *td = &tif->tif_dir;
	uint32 nstrips;

+   if( td->td_nstrips )
+       return td->td_nstrips;
+
	    nstrips = (td->td_rowsperstrip == (uint32) -1 ? 1 :
	        TIFFhowmany_32(td->td_imagelength, td->td_rowsperstrip));
	    if (td->td_planarconfig == PLANARCONFIG_SEPARATE)
\end{lstlisting}

As shown in Listing~\ref{lst:gnubug-26545}, the off-by-one error occurs because the loop condition uses \texttt{\small <} instead of \texttt{\small <=}. When the \texttt{\small size} is an even number, the value of \texttt{\small i} may exactly equal \texttt{\small size/2}, so with the previous \texttt{\small <} condition, the loop may exit early and fail to handle half the size in certain cases. By changing it to \texttt{\small{\textit{<=}}}, it ensures that the loop will execute when \texttt{\small{i}} equals \texttt{\small{size/2}}, fixing the vulnerability where the buffer wasn't fully filled under specific inputs. This vulnerability was not repaired by any non-learning based methods because lack of correct constraint, whereas learning-based methods were able to adaptively infer or generate the necessary conditions for repair for such vulnerabilities without complex dependencies. Correspondingly, as shown in Listing~\ref{lst:cve-2016-10094}, non-learning methods extracted constraints correctly and both methods could lead to a correct repair.

In Listing~\ref{lst:cve-2016-9273}, the vulnerability fix is a single-hunk fix and only has two lines of code addition. However, it cannot be repaired by both learning and non-learning methods. The root of the vulnerability lies in the reliance on the cached value of \texttt{td->td\_nstrips}, which was originally computed to avoid redundant calculations. This value is calculated when the strip count is first needed, based on the image's length and the rows per strip. However, if the structure of the image changes afterward in another function, the cached value becomes outdated, leading to inconsistencies. The vulnerability emerges because there is no mechanism to ensure that the cached \texttt{td->td\_nstrips} value is updated when the underlying image structure changes. For learning-based methods, it has no context, no information about the \texttt{td} member, and unclear about the root cause of this vulnerability, so it's hard to repair. For non-learning based methods, wrong constraints, and wrong localization lead to no successul repair. As analyzed in this case, although the change is 2 lines of code, it still has a complex logic. 
% For more case studies, please refer to our \href{https://avr-sok.netlify.app/cases/}{website}. 

\subsection{Future Directions}
\label{app:future}
\subject{D8. Vulnerability Repair for Binaries.} \newtext{As we discussed in Section~\ref{sec:taxonomy_apr}, most SPG focuses on source code-level fixes, which poses several limitations: (1) It requires access to the original build environment, which may be unavailable for older applications. (2) Developers may delay patching vulnerabilities in third-party libraries~\cite{duck2020binary}. (3) Managing complex configurations and build options of open-source software is challenging~\cite{duan2019automating}. To implement binary-level vulnerability fixes, we advocate integrate current repair methods with binary-rewriting or CodeLLMs for binary-level AVR.}

\subject{D9. AVR Tools Usability.} AVR can help reduce the workload of human. But how to better guide the use of the AVR tool effectively is still a challenging issue.  Currently, most AVR tools are command-line based, which can present usability barriers for developers. Future research could also explore the AVR tools usability and develope usable AVR tools, previous work~\citep{smith2020can, tripp2014aletheia} has stressed the usability of static security analysis tools but never for AVR tools.

%-------------------------------------------------------------------------------
% Appendix
%-------------------------------------------------------------------------------
% \input{tex/appendix.tex}
%-------------------------------------------------------------------------------

%TC:endignore
%-------------------------------------------------------------------------------
\end{document}